\newif\ifextended
\newif\ifcomments

\extendedtrue

\commentsfalse

\documentclass[acmsmall,nonacm]{acmart} 

\setcopyright{rightsretained}
\acmDOI{10.1145/3839462}
\acmYear{2026}
\acmJournal{PACMPL}
\acmVolume{10}
\acmNumber{OOPSLA2}
\acmArticle{330}
\acmMonth{10}

\setcitestyle{nosort}

\usepackage{booktabs}   
\usepackage{subcaption} 

\usepackage[utf8]{inputenc}
\usepackage[english]{babel}
\usepackage{xparse}
\usepackage{xspace}
\usepackage{thmtools}

\ifcomments
\usepackage[draft]{fixme} 
\else
\usepackage[final]{fixme} 
\fi

\usepackage{xcolor}
\usepackage{mathpartir}
\usepackage[]{microtype}
\usepackage{multirow}
\usepackage{makecell}
\usepackage{marvosym}
\usepackage{wasysym}
\usepackage{pifont}
\usepackage{mathtools}
\usepackage{stmaryrd}
\usepackage{scalerel}
\usepackage{tensor}
\usepackage{xifthen}
\usepackage{iris}
\usepackage{pftools}
\usepackage{soul}
\usepackage{tikz}
\usepackage{threeparttable}
\usepackage{fontawesome}
\usepackage{enumitem}
\usetikzlibrary{calc, shapes, arrows, automata, patterns}

\makeatletter
\addto\extrasenglish{%
  \renewcommand*\chapterautorefname{\S\@gobble}
  \renewcommand*\sectionautorefname{\S\@gobble}
  \renewcommand*\subsectionautorefname{\S\@gobble}
  }
\makeatother
\newcommand*{\myeqref}[2][]{%
  \hyperref[{#2}]{#1(\ref*{#2})}%
}

\newcommand*{\lineref}[1]{\hyperref[#1]{line~\ref*{#1}}}
\newcommand*{\linerangeref}[2]{\hyperref[#1]{lines~\ref*{#1}}\hyperref[#2]{-\ref*{#2}}}
\newcommand{\mypageref}[1]{\hyperref[#1]{page~\pageref*{#1}}}

\newcommand{\appendixref}[2]{\ifextended{Appendix~\ref{#1}}\else{the extended version~\cite[\S#2]{extendedversion}}\fi}

\declaretheorem[name=Definition,style=definition]{definition}
\declaretheorem[name=Theorem,sibling=definition]{theorem}

\DeclareUnicodeCharacter{2264}{$\le$}
\DeclareUnicodeCharacter{2208}{$\in$}
\DeclareUnicodeCharacter{2203}{$\exists$}
\DeclareUnicodeCharacter{25C1}{$\triangleleft$}
\DeclareUnicodeCharacter{2097}{${}_l$}
\DeclareUnicodeCharacter{2260}{$\neq$}
\DeclareUnicodeCharacter{2205}{$\emptyset$}
\DeclareUnicodeCharacter{222A}{$\cup$}
\DeclareUnicodeCharacter{228E}{$\uplus$}
\DeclareUnicodeCharacter{2200}{$\forall$}
\DeclareUnicodeCharacter{25A1}{$\always$}
\DeclareUnicodeCharacter{03B3}{$\gamma$}

\definecolor{airforceblue}{rgb}{0.36, 0.54, 0.66}
\definecolor{brickred}{rgb}{0.9, 0.55, 0}
\definecolor{ao}{rgb}{0.0, 0.0, 1.0}
\definecolor{cobalt}{rgb}{0.0, 0.28, 0.67}
\definecolor{darkergreen}{rgb}{0,0.7,0.3}
\definecolor{magenta}{rgb}{1.0,0.0,1.0}

\FXRegisterAuthor{td}{atd}{TD}
\FXRegisterAuthor{ms}{ams}{MS}
\FXRegisterAuthor{gp}{agp}{GP}
\FXRegisterAuthor{as}{aas}{AS}
\FXRegisterAuthor{pm}{apm}{PM}

\makeatletter
\providecommand*{\Dashv}{%
  \mathrel{%
    \mathpalette\@Dashv\vDash
  }%
}
\newcommand*{\@Dashv}[2]{%
  \reflectbox{$\m@th#1#2$}%
}
\makeatother
\makeatletter 
\def\arcr{\@arraycr}
\makeatother

\newcommand\ie{\emph{i.e.}, }
\newcommand\eg{\emph{e.g.}, }

\makeatletter
\def\@parfont{\bfseries\itshape}
\makeatother

\usepackage{tikz}
\usetikzlibrary{calc}
\usetikzlibrary{shapes.geometric, arrows}

\tikzstyle{question} = [rectangle,
minimum width=3cm,
minimum height=1cm,
text centered,
text width=3cm,
draw=black,
fill=orange!30]

\tikzstyle{rule} = [rectangle, rounded corners,
minimum width=1.5cm,
minimum height=1cm,
text centered,
draw=black,
fill=blue!20]

\tikzstyle{arrow} = [thick,->,>=stealth]

\usepackage{listings, silver, front-end}
\usepackage{color}
\usepackage{soul}
\usepackage{makecell}
\usepackage{multirow}
\usepackage{dirtree}
\usepackage{xspace}
\usepackage{float}

\usepackage{caption}
\usepackage{subcaption}

\usepackage{hyperref}
\usepackage{breakurl}

\usepackage{wrapfig}

\definecolor{light-gray}{gray}{0.9}
\definecolor{darkgreen}{rgb}{0.0,0.6,0.0}

\usepackage{tikz}

\usepackage[scaled=0.85]{beramono}
\usepackage[T1]{fontenc}

\newcommand\code[1]{{\texttt{#1}}}
\newcommand\secref[1]{\autoref{#1}}
\newcommand\appref[1]{\PackageError{vipermacros}{use appendixref instead of appref}{}}
\newcommand\defref[1]{\hyperref[#1]{Def.~\ref*{#1}}}
\newcommand\notref[1]{\hyperref[#1]{Not.~\ref*{#1}}}
\newcommand\lstref[1]{\hyperref[#1]{List.~\ref*{#1}}}
\newcommand\tabref[1]{\hyperref[#1]{Tab.~\ref*{#1}}}
\newcommand\figref[1]{\hyperref[#1]{Fig.~\ref*{#1}}}
\newcommand\thmref[1]{\hyperref[#1]{Thm.~\ref*{#1}}}
\newcommand\lemref[1]{\hyperref[#1]{Lemma~\ref*{#1}}}
\newcommand\propref[1]{\hyperref[#1]{Prop.~\ref*{#1}}}
\newcommand\exref[1]{\hyperref[#1]{Ex.~\ref*{#1}}}
\newcommand\reqref[1]{\hyperref[#1]{Req.~\ref*{#1}}}
\newcommand\simpreqref[1]{\hyperref[#1]{Simp.~Req.~\ref*{#1}}}

\newcommand{\wrt}{{{w.r.t.\@}}}

\usepackage[normalem]{ulem}

\ifcomments\else
\def\nocolour{ }
\fi

\newcommand\todo[1]{\ifdefined\nocolour{}\else{\textcolor{red}{TODO: #1}}\fi}

\newcounter{colorctxdepth}
\newcommand{\colorcontext}[2]{%
  \addtocounter{colorctxdepth}{1}%
  \textcolor{#1}{#2}%
  \addtocounter{colorctxdepth}{-1}%
}
\DeclareRobustCommand{\revision}[1]{\ifdefined\revisioncolor{\colorcontext{darkgreen}{#1}}\else#1\fi}

 \newcommand{\peter}[1]{\ifdefined\nocolour{#1}\else{\color{brickred}{#1}}\fi}

\newcommand{\hongyi}[1]{\ifdefined\nocolour{{#1}}\else{\color{brown}{#1}}\fi}

\newcommand{\isabelle}{Isabelle/HOL}

\definecolor{darkred}{rgb}{0.55, 0.0, 0.0}

\usepackage{mathtools}
\usepackage{comment}

\usepackage{prooftree}

\makeatletter

\newskip \point

\def \premisespacing{\quad}

\def \RulePremisesNewlineMore[#1]#2.#3#4{\@ifnextchar\bgroup{\RulePremisesNewlineMore[#1]{#2}.{#3\premisespacing#4}}{\@ifnextchar.{\RulePremisesNewline[#1]{{\begin{array}{c}#2\\#3\premisespacing#4\end{array}}}}{\RuleMultiPremise[#1]{{\begin{array}{c}#2\\#3\end{array}}}{#4}}}}

\def \RulePremisesNewline[#1]#2.#3{\@ifnextchar\bgroup{\RulePremisesNewlineMore[#1]{#2}.{#3}}{\@ifnextchar.{\RulePremisesNewline[#1]{{\begin{array}{c}#2\\#3\end{array}}}}{\RuleMultiPremise[#1]{#2}{#3}}}}

\def \RuleMultiPremise[#1]#2#3{\@ifnextchar\bgroup{\RuleMultiPremise[#1]{#2\premisespacing#3}}{\@ifnextchar.{\RulePremisesNewline[#1]{#2\premisespacing#3}}{\prooftree #2\justifies#3 \using{#1}\endprooftree}}}

\def \RuleWithName[#1]#2{\@ifnextchar\bgroup {\RuleMultiPremise[#1]{#2}}{\@ifnextchar.{\RulePremisesNewline[#1]{#2}}{\prooftree \justifies #2 \using{#1} \endprooftree}}}

\def \RuleWithInfo[#1]{\@ifnextchar[{\RuleWithNameAndCondition[#1]}{\RuleWithName[(#1)]}}

\def \RuleWithNameAndCondition[#1][#2]{\RuleWithName[(#1)^{#2}]}

\def \Inf{\proofrulebaseline=2ex \abovedisplayskip12\point\belowdisplayskip12\point \abovedisplayshortskip8\point\belowdisplayshortskip8\point \@ifnextchar[{\RuleWithInfo}{\RuleWithName[ ]}}

\makeatother

\usepackage{amsthm}

\newtheorem{lemma}{Lemma}

\newtheorem{simpreq}{Simplified Requirement}
\newtheorem{requirement}{Requirement}
\newcounter{restoreReqCnt}
\makeatletter
\newenvironment{grequirement}[2][]{%
  \def\@tempa{#1}%
  \setcounter{restoreReqCnt}{\value{requirement}}
  \ifx\@tempa\@empty
    \begin{requirement}%
  \else
    \begin{requirement}[#1]%
  \fi
}{%
  \end{requirement}%
  \setcounter{requirement}{\value{restoreReqCnt}}
}
\makeatother


\usepackage{xcolor}

\usepackage{esvect}

\renewcommand*{\proofname}{Proof sketch}

\usepackage{pifont}

\definecolor{royalblue}{rgb}{0.25, 0.41, 0.88}

\definecolor{somebrown}{rgb}{0.8, 0.35, 0.1}

\newcommand{\emp}{\ensuremath{ \mathit{emp} }}

\newcommand*{\pointsto}[2]{\ensuremath{{#1} \mapsto {#2}}}

\newcommand{\axiomSem}[4][\Delta]{\ensuremath{#1 \vdash\allowbreak [#2] \,\allowbreak #3 \,\allowbreak [#4]}}
\newcommand{\stable}[1]{\textsf{stable}(#1)}
\newcommand{\stabilize}[1]{\textsf{stabilize}(#1)}
\newcommand{\stableNa}{\textsf{stable}}
\newcommand{\stabilizeNa}{\textsf{stabilize}}
\renewcommand{\unit}[1]{\textsf{unit}({#1})}
\newcommand{\unitNa}{\textsf{unit}}

\newcommand{\selfFraming}[1]{\textsf{selfFraming}(#1)}
\newcommand{\framedBy}[2]{\textsf{frames}(#1, #2)}

\newcommand{\core}[1]{|#1|}

\newcommand{\Variable}{\mathsf{Var}}

\newcommand{\vfalse}{\vkeyword{false}}
\newcommand{\vtrue}{\vkeyword{true}}

\usepackage{stmaryrd}
\usepackage{tabularx}
\newcolumntype{Y}{>{\centering\arraybackslash}X}

\allowdisplaybreaks

\begin{document}

\ifextended
\title{Sound State Encodings in Translational Separation Logic Verifiers (Extended Version)}
\else
\title{Sound State Encodings in Translational Separation Logic Verifiers}
\fi

\author{Hongyi Ling}
\orcid{0000-0003-3383-3482}
\affiliation{%
  \institution{ETH Zurich}
  \city{Zurich}
  \country{Switzerland}
}
\email{hongyi.ling@inf.ethz.ch}

\author{Thibault Dardinier}
\orcid{0000-0003-2719-4856}
\affiliation{%
  \institution{EPFL}
  \city{Lausanne}
  \country{Switzerland}
}
\email{thibault.dardinier@epfl.ch}

\author{Ellen Arlt}
\orcid{0009-0003-4620-3137}
\affiliation{%
  \institution{MPI-SWS}
  \city{Kaiserslautern}
  \country{Germany}
}
\email{earlt@mpi-sws.org}

\author{Peter Müller}
\orcid{0000-0001-7001-2566}
\affiliation{%
  \institution{ETH Zurich}
  \city{Zurich}
  \country{Switzerland}
}
\email{peter.mueller@inf.ethz.ch}

\begin{abstract}
Automated program verifiers are often organized into a front-end, which encodes an input program into an intermediate verification language (IVL), and a back-end, which proves that the IVL program is correct.
Soundness of such translational verifiers requires that the back-end verification is sound and that correctness of the IVL program implies correctness of the input program. 
Existing formalizations for translational verifiers based on separation logic target the former, but support the latter only under the strong assumption that there exists a separation logic for the input program with the same state model as the IVL.
This assumption is unrealistic in practice, especially since the state model also defines the supported separation logic resources.

We present the first formal framework for proving the soundness of translational separation logic verifiers with non-trivial state encodings.
To be applicable to various front-ends and IVLs, our framework only assumes the existence of a homomorphic encoding relation between the front-end and IVL state models.
At the core of our framework is a novel condition, backward satisfiability, which is crucial to guarantee the soundness of the front-end translation.
We formalize our framework for front-end verifiers based on concurrent separation logic and separation logic IVLs, such as Raven, VeriFast, and Viper.
We demonstrate its expressiveness by proving soundness for three common state encodings.
Our framework and all proofs are formalized in \isabelle{}.
\end{abstract}



\maketitle

\section{Introduction}
\label{sec:intro}

Many modern program verification tools are \emph{translational verifiers}, \ie they split the verification into \emph{front-ends}, which translate the input program with its specification into an \emph{intermediate verification language} (IVL), and \emph{back-ends}, which verify the resulting IVL program. This split enables the reuse of back-end verifiers across multiple front-ends, thus dramatically reducing the effort of developing verifiers. Examples of translational verifiers include Civl~\cite{kraglCivlVerifier2021} and Dafny~\cite{leinoDafnyAutomaticProgram2010} based on the Boogie IVL~\cite{leinoThisBoogie22008}, Gillian for C and Javascript~\cite{maksimovicGillianPartII2021} and Rust~\cite{ayounHybridApproachSemiautomated2025} based on GIL~\cite{fragoso2020gillian},  Prusti~\cite{astrauskasLeveragingRustTypes2019}, Nagini~\cite{eilersNaginiStaticVerifier2018}, and Gobra~\cite{wolfGobraModularSpecification2021} based on Viper~\cite{mullerViperVerificationInfrastructure2016}, as well as 
Frama-C~\cite{CuoqKKPSY12} and Creusot~\cite{DenisJM22} based on WhyML~\cite{FilliatreP13}.
To support heap-manipulating and concurrent programs, many existing translational verifiers and their IVLs (GIL, Raven~\cite{GuptaPW25}, VeriFast\footnote{VeriFast does not have a human-readable IVL, but uses a uniform internal representation and back-end for multiple input languages (C, Java, Rust).}~\cite{JacobsSPVPP11}, and Viper) are based on separation logic (SL)~\cite{reynoldsSeparationLogicLogic2002}.

Soundness of a translational verifier requires (1)~soundness of the back-end verifier and (2)~soundness of the front-end translation, that is, that the successful verification of the IVL program implies correctness of the input program. While several existing works target back-end soundness for various IVLs~\cite{loow2025compositional,THSem,ParthasarathyMuellerSummers21,Cohen24,Herms13}, the state of the art does not offer practical techniques to prove soundness of realistic front-ends based on SL.

\begin{figure*}
\newcommand{\highl}[2]{\colorbox{#1}{$\displaystyle #2$}}
\newcommand{\preA}{\ensuremath{\highl{yellow!30}{\code{true}}}}
\newcommand{\preB}{\ensuremath{\highl{blue!20}{x.f \mapsto \_}}}
\newcommand{\preBIVL}{\ensuremath{\highl{blue!20}{\existsntp{b}{x.\addedf{} \mapsto b * (b \Rightarrow x.\ivlf{} \mapsto \_)}}}}
\newcommand{\postA}{\ensuremath{\highl{green!25}{y = v}}}
\newcommand{\postB}{\ensuremath{\highl{red!20}{\existsntp{v'}{x.f \mapsto v' * v' = v}}}}
\newcommand{\postBIVL}{\ensuremath{\highl{red!20}{\existsntp{v'}{x.\addedf{} \mapsto \vtrue{} * x.\ivlf{} \mapsto v' * v' = v}}}}
\footnotesize
\begin{subfigure}{\textwidth}
\begin{frontend}[mathescape]
def main(v: Int):
   x := alloc(f)
// {$\color{darkgreen}\preA$}      {$\color{darkgreen}\preB$}
   y := v   ||   x.f := v
// {$\color{darkgreen}\postA$}      {$\color{darkgreen}\postB$}
   assert y = x.f
\end{frontend}
\caption{Python-like front-end program, with concurrency and lazy field creation.
The comments before and after the parallel composition represent the pre- and postconditions of the two threads, respectively.
Here $x.f \mapsto \_$ represents the ownership of $x.f$ \emph{without} any guarantee that the field \code{f} has been created yet,
while $\existsntp{v'}{x.f \mapsto v'}$ represents the ownership of $x.f$ \emph{with} the guarantee that it has been created.}
\label{fig:dfc-ex:front-end}
\end{subfigure}

\vspace{10pt}

\begin{subfigure}{\textwidth}
\begin{minipage}{0.5\textwidth}
\begin{viper}[mathescape]
method main(v: Int) {
   // x := alloc(f)
   havoc x
   inhale $x.\addedf{} \mapsto \vfalse{}$

   // parallel composition
   exhale $\preA$
   exhale $\preBIVL$
   havoc y
   inhale $\postA$
   inhale $\postBIVL$

   // assert y = x.f
   assert $\existsntp{v'}{x.\ivlf{} \mapsto v' * y = v'}$
}

$\phantom{x}$
\end{viper}
\end{minipage}
\begin{minipage}{0.5\textwidth}
\begin{viper}[mathescape]
method leftThread(v, y: Int) {
   inhale $\preA$
   y := v
   exhale $\postA$
}

method rightThread(v: Int, x: Ref) {
   inhale $\preBIVL$

   // x.f := v
   if ($\lnot$x.$\addedf{}$)
      inhale $x.\ivlf{} \mapsto \_$
   x.$\addedf{}$ := true
   x.$\ivlf{}$ := v

   exhale $\postBIVL$
}
\end{viper}
\end{minipage}
\caption{Example translation of the front-end program into the Viper IVL, inspired by Nagini.
The highlighted assertions represent the translations of the corresponding pre- and postconditions from \figref{fig:dfc-ex:front-end}.}
\label{fig:dfc-ex:encoding}
\end{subfigure}

\caption{An example front-end translation.}
\label{fig:dfc-ex}
\end{figure*}

\subsection{Challenge: Large Semantic Gap}

The main challenge in establishing the soundness of a front-end translation into an SL-based IVL is to bridge the large semantic gap between the semantics and proof rules of the front-end programming language and the IVL\@.
As an example, consider the program in \figref{fig:dfc-ex:front-end}.
This program allocates a new object with field \code{f},
then concurrently assigns the value of the argument \code{v} to local variable \code{y} and field \code{x.f},
before asserting that \code{y} and \code{x.f} are equal.
Like in Python, the language used here adds
the field \code{f} lazily to the object when it is first written:
Attempting to read a field that has not yet been added results in a runtime error.
To verify this program, a translation into the IVL must check
that all fields that are read were previously assigned,
in addition to ensuring the absence of memory errors and data races, and that the assertion at the end of method \code{main} holds.

\paragraph{Concurrency}
Our IVL translation, inspired by the Viper-based Python verifier Nagini~\cite{eilersNaginiStaticVerifier2018}, is shown on \figref{fig:dfc-ex:encoding}.
Since the IVL does not support concurrency directly, we translate it using concurrent separation logic (CSL)~\cite{ohearnResourcesConcurrencyLocal2004} rules:
The two front-end threads with their own pre- and postconditions are translated into and verified as separate IVL methods \vcode{leftThread} and \vcode{rightThread}. In each of them, the translated precondition is added to the initially-empty IVL state via an \vcode{inhale} (also called \emph{produce}) operation, the translated thread body is executed, and the translated postcondition is checked (and removed from the final state) via \vcode{exhale} (also called \emph{consume}).
In the main method, the concurrent part is abstracted to an exhale-havoc-inhale sequence that first removes the resources transferred to each thread via \vcode{exhale}s, then reflects the side effects of the threads on the local variable \code{y} by assigning a non-deterministic value (\vcode{havoc}), and finally adds the resources given back by each thread via \vcode{inhale}s.

\paragraph{Lazy field creation}
As our target IVL has static field declarations (but does not support lazy field creation directly),
our encoding needs to reflect for each field whether it has been added. To this end, the encoding represents each field $f$ of the front-end program with two statically-declared fields in the IVL: the field $\ivlf{}$ to store the value of the front-end field, and an additional boolean field $\addedf{}$ that indicates whether field $f$ has been added to the object.
As shown in the example:
(1)~Allocating the object $x$ (as in the translation of \fecode{x := alloc(f)}) produces the SL resource $x.\addedf{} \mapsto \vfalse$, which provides the permission to add the field later, but does \emph{not} generate any ownership to $\ivlf{}$.
(2)~Assigning to $x.f$ (as in the translation of \fecode{x.f := v} in the right thread) requires $x.\addedf{}\mapsto \vfalse$ (if the field has not been added before) or $x.\ivlf{}\mapsto \_$ (if it has). In the former case, the resource $x.\ivlf{} \mapsto \_$ is produced lazily. In either case, $x.\ivlf{} \mapsto v$ is held after the assignment, thereby enabling subsequent reads.
(3)~Reading $x.f$ (as in the translation of \fecode{assert y = x.f}) requires $x.\ivlf{}\mapsto \_$, as usual.

\medskip
This example demonstrates that the gap between front-end programs and their IVL encoding is substantial, affecting the state representation (here, an extra IVL field per front-end field), the operational execution (\eg concurrency in the front-end program vs.\ multiple sequential methods on the IVL), and proof rules (\eg CSL for the front-end vs.\ \code{inhale} and \code{exhale} primitives in the IVL). A soundness proof for a front-end translation needs to bridge this gap to show that verification of the IVL program implies the correctness of the front-end program.

\subsection{State of the Art}

\begin{figure}
   \centering
   \tikzstyle{every node}=[font=\footnotesize]
   \definecolor{yellowboxcolor}{HTML}{ffffb3}
   \tikzstyle{block}=[inner sep=0.2em, minimum height=0.8cm, draw, align=center, preaction={fill,white}]
   \begin{tikzpicture}[
    arrow/.style={
        ->,
        >=Stealth,
        thick
    }
]

\node[block, fill=yellowboxcolor] (B1) at (0, 0) {Front-end program correct \\\wrt{} operational semantics};
\node[block, fill=yellowboxcolor] (B2) at (2,1.8) {Valid front-end\\SL proof};
\node[block, fill=yellowboxcolor] (B3) at (7.8,1.8) {Translation correct \wrt{}\\IVL \emph{axiomatic} semantics};
\node[block, fill=yellowboxcolor] (B4) at (10.6,0) {Translation correct \wrt{}\\IVL \emph{operational} semantics};

\draw[arrow] (B4.north)
   -- ($(B4.north |- B3.east)$)
   node[midway, left, yshift=-0.15cm, font=\footnotesize, align=center]
   {(1) Equivalence of\\the two semantics~\cite{dardinierFormalFoundationsTranslational2025}}
   -- (B3);

\draw[arrow] (B3.west) -- (B2.east)
   node[midway, above, font=\footnotesize, align=center]
   {(2) Proof reconstruction};

\draw[arrow] (B2)
   -- ($(B1.north |- B2.west)$)
   -- (B1)
   node[midway, right, yshift=-0.15cm, font=\footnotesize, align=center]
   {(3) Soundness of front-end SL};

\draw[arrow, dashed] (B4.west)
   -- (B1.east)
   node[midway, above, font=\footnotesize, align=center]
   {Soundness of the front-end translation};

\end{tikzpicture}
  \caption{Decomposing the soundness proof via a front-end SL. The dashed arrow represents the goal and the three solid arrows represent the decomposition. Existing work does the decomposition, but assumes that the two SLs have the same state model, which is unrealistic in practice. Our work lifts this key restriction, and enables full soundness proofs for front-ends with different state models to their IVLs.}
  \label{fig:decomposition}
\end{figure}

As explained above, establishing the soundness of a front-end translation into an SL-based IVL, for example for the translation depicted in \figref{fig:dfc-ex},
requires bridging the large semantic gap between the semantics and proof rules of the front-end programming language and the IVL\@.
\citet{dardinierFormalFoundationsTranslational2025} show that this gap can be reduced to proving that
the correctness of the IVL encoding (with respect to an \emph{axiomatic} semantics)
implies the existence of a valid proof in a suitable \emph{program logic} for the front-end programming language,
as depicted in \figref{fig:decomposition}.
This decomposition leverages the existence of sound separation logics for many advanced front-end features
(\eg different kinds of concurrency \cite{ohearnResourcesConcurrencyLocal2004, gotsman2007local, dinsdale-youngConcurrentAbstractPredicates2010, VafeiadisN13, jungIrisMonoidsInvariants2015}, initialization semantics~\cite{PereiraBFE025}, non-local control flow~\cite{krebbers2013separation}, etc.) to prove the soundness of front-end translations.
For example, Dardinier et al.\  show how to prove the soundness of the concurrency translation from \figref{fig:dfc-ex}.
However, their
framework requires the front-end SL and the IVL to use the same state model, which is unrealistic in practice.

In practice, most front-ends use a state model that is vastly different from the fixed state model of the IVL\@.
One major discrepancy is how memory and data are represented. For example, C's memory regions or Java's arrays might be encoded into an object-based memory model in an IVL, which does not natively support a notion of consecutive addresses. Moreover, our example from \figref{fig:dfc-ex:front-end} encodes each Python field by a pair of fields to track which fields have been added to an object.

Important discrepancies also arise for the information tracked as part of the verification state, such as the separation logic resources held in a state. For example, front-end states may include designated resources to execute a method call~\cite{jacobs2015modular} or perform an I/O operation~\cite{penninckxSoundModularCompositional2015,sprenger_igloo_2020}, which need to be encoded into standard SL predicates in the IVL\@.
Moreover, front-end and IVL may use a different permission model. For example, Viper~\cite{mullerViperVerificationInfrastructure2016}, VeriFast~\cite{JacobsSPVPP11}, and Raven%
\footnote{Raven employs user-definable resource algebras for its ghost resources, but still uses fractional permissions for its regular program states.}~\cite{GuptaPW25}
all build in fractional permissions~\cite{boylandCheckingInterferenceFractional2003,bornatPermissionAccountingSeparation2005}
but have front-ends that use binary%
\footnote{That is, one either owns a full permission or no permission at all.}
permissions (\eg the encoding of I/O tokens in VeriFast~\cite{penninckxSoundModularCompositional2015} or Viper~\cite{sprenger_igloo_2020,pereira_protocols_2025}),
counting permissions~\cite{bornatPermissionAccountingSeparation2005} (\eg in Gobra~\cite{wolfGobraModularSpecification2021} and the encoding of relaxed separation logic~\cite{VafeiadisN13} in Viper~\cite{SummersM20, mullerViperVerificationInfrastructure2016}), or duplicable permissions (\eg duplicable invariants over immutable shared data~\cite{PereiraBFE025}).


Due to these discrepancies, none of these encodings can be proven sound by the existing framework by Dardinier et al.

\subsection{This Work}

We develop the first formal framework for proving the soundness of front-end translations that use different state models from their IVLs, thus lifting the key restriction of  existing work  and enabling soundness proofs for practical front-end translations.
To be compatible with a wide range of front-end separation logics and IVLs, our framework simply assumes the existence of a homomorphic encoding relation between front-end and IVL states.
We identify a novel condition on this encoding relation, which we call \emph{backward satisfiability}, that is crucial to guarantee the soundness of the front-end translation.
To show the practicality of our approach, we develop a formal framework to prove the soundness of front-end translations based on concurrent separation logic (CSL)~\cite{ohearnResourcesConcurrencyLocal2004}, and instantiate it with three different front-end translations inspired by existing Viper front-ends, including the one in \figref{fig:dfc-ex}.
We also show that backward satisfiability is essential for soundness, by uncovering several subtle sources of unsoundness in state encodings that affect existing verifiers, but have not been described before.
\revision{More broadly, we show that violations of our framework's requirements underlie several previously-reported soundness bugs in verifiers such as Prusti, Nagini, and VerCors~\cite{blomVerCorsToolSet2017}.}

\revision{
In addition to proving the soundness of existing front-ends, our framework also guides the development of new sound front-end translations.
Developing a front-end typically involves three steps: choosing a suitable IVL, designing the state encoding relation between front-end and IVL states, and defining the translations of front-end expressions, assertions, and statements. Our framework's requirements support each of these steps in two ways. First, they provide design guidance even before any proof is attempted: because a sound design must satisfy them, they rule out large families of unsound choices and narrow the design space. Second, they enable incremental verification: once a design decision is made, the corresponding requirement is a self-contained proof obligation that can be discharged immediately, allowing to catch an unsound choice early rather than after the whole translation is built.
}

In summary, we make the following technical contributions:
\begin{enumerate}
   \item We identify a set of sufficient conditions, including the novel \emph{backward satisfiability}, that guarantee the soundness of state encodings (\secref{sec:key} and \secref{sec:framework}).
   \item We identify several subtle sources of unsoundness in state encodings that affect existing verifiers, but have not been described before (\secref{sec:key}).
   \item We develop a formal framework for proving the soundness of front-end translations. It is parametric in the front-end state model, expressions, assertions, statements, and logic, as well as their translations into the IVL\@. It is applicable to the translations used in practical SL-verifiers for both sequential and concurrent programs (\secref{sec:framework}).
   \item We demonstrate the expressiveness of our framework by instantiating it for three practical front-end translations inspired by existing verifiers (\secref{sec:instantiations}).
   \item The framework including all proofs is formalized in \isabelle{}.
\end{enumerate}

Our framework, including all formalizations and proofs, is mechanized in \isabelle{}~\cite{nipkowIsabelleHOLProof2002} and the mechanization is publicly available~\cite{ling_artifact}.

\section{Overview: Homomorphism, Spurious Splittability, and Backward Satisfiability}
\label{sec:key}

In this section, we present key properties that a front-end translation must satisfy to be sound.
We assume the existence of a state encoding relation between front-end and IVL states, which is general enough to capture many practical front-end translations (\secref{subsec:relation}).
We explain why this relation should be homomorphic, and show how violating this requirement can lead to unsoundness
(\secref{subsec:homomorphism}).
We then describe the key problem of \emph{spurious splittability}, where the IVL state resulting from encoding a front-end state is split into two IVL states that have no front-end counterparts (\secref{subsec:spurious-splittability}). We explain how spurious splittability may lead to unsoundness and introduce our solution, \emph{backward satisfiability}, to prevent this (\secref{subsec:backward-satisfiability}).
Finally, we show examples where spurious splittability, and the absence of backward satisfiability, lead to unsoundness
(\secref{subsec:unsoundness}).

\begin{figure*}
  \centering
  \footnotesize
\begin{tikzpicture}

  \node (ivl) at (0, 0) {};
  \node (fe) at (0, 3.1) {};
  \pgfmathsetmacro{\fey}{1.1};
  \pgfmathsetmacro{\fex}{\fey * 2.1};
  \pgfmathsetmacro{\ivly}{1.8};
  \pgfmathsetmacro{\ivlx}{\ivly * 2.1};
  \pgfmathsetmacro{\disivlx}{-\ivlx * 0.9};
  \pgfmathsetmacro{\disivly}{-\ivly * 0.8};
  \pgfmathsetmacro{\disfex}{-\fex * 1.3};
  \pgfmathsetmacro{\disfey}{\fey * 0.75};
  \pgfmathsetmacro{\arrowendy}{0.1};
  \pgfmathsetmacro{\wx}{1.3};
  \pgfmathsetmacro{\wy}{-0.2};
  \pgfmathsetmacro{\Wfx}{-2.1};
  \pgfmathsetmacro{\Wfy}{1.3};
  \pgfmathsetmacro{\Wax}{0.3};
  \pgfmathsetmacro{\Way}{-1.45};
  \pgfmathsetmacro{\wfx}{0.25};
  \pgfmathsetmacro{\wfy}{0.8};
  \pgfmathsetmacro{\wax}{-0.6};
  \pgfmathsetmacro{\way}{-0.4};

  \draw[thick] (ivl) ellipse (\ivlx cm and \ivly cm);
  \node at ($(ivl)+(\disivlx, \disivly)$) {\textbf{IVL states}};

  \draw[thick, dashed] (ivl) ellipse (\fex cm and \fey cm);

  \draw[thick] (fe) ellipse (\fex cm and \fey cm);
  \node at ($(fe)+(\disfex, \disfey)$) {\textbf{Front-end states}};

  \draw[thick, dashed, ->, >=Stealth]
    ($(fe)+(-\fex, 0)$) -- ($(ivl)+(-\fex, \arrowendy)$);

  \draw[thick, dashed, ->, >=Stealth]
    ($(fe)+(\fex, 0)$) -- node[right] {$\trs{\cdot}$} ($(ivl)+(\fex, \arrowendy)$);

  \node[circle, fill=black, inner sep=1.5pt, label={[xshift=-0.05cm, yshift=0.1cm]right: $\omega$}] (wb) at ($(fe)+(\wx, \wy)$) {};

  \node[circle, fill=black, inner sep=1.5pt, label={right: $\trs{\omega}$}] (trwb) at ($(ivl)+(\wx, \wy)$) {};

  \draw[thick, ->, >=Stealth] (wb) -- ($(trwb)+(0, \arrowendy)$);

  \node[circle, fill=red, inner sep=1.5pt, label={[red, yshift=0.05cm]below:$\Omega_F$}] (W1) at (\Wfx,  \Wfy) {};
  \node[circle, fill=red, inner sep=1.5pt, label={[red]right:$\Omega_A \vDash \tra{A}$}] (W) at (\Wax, \Way) {};

  \draw[red, thick, ->, >=Stealth] (trwb) -- (W1);
  \draw[red, thick, ->, >=Stealth] (trwb) -- (W);

  \node[circle, fill=blue!50, inner sep=1.5pt, label={[blue!50, yshift=0.1cm]right:$\omega_F$}] (w1) at ($(fe)+(\wfx, \wfy)$) {};
  \node[circle, fill=blue!50, inner sep=1.5pt, label={[blue!50, yshift=-0.05cm, xshift=0.1cm]above:$\omega_A \vDash A$}] (w) at ($(fe)+(\wax, \way)$) {};

  \draw[black!50, dotted, thick, ->, >=Stealth] (wb) -- (w1);
  \draw[black!50, dotted, thick, ->, >=Stealth] (wb) -- (w);

  \node[circle, fill=blue, inner sep=1.5pt, label={[blue]right:$\trs{\omega_F}$}] (w1i) at ($(ivl)+(\wfx, \wfy)$) {};
  \node[circle, fill=blue, inner sep=1.5pt, label={[blue]left:$\trs{\omega_A}$}] (wi) at ($(ivl)+(\wax, \way)$) {};

  \draw[black!50, dotted, thick, ->, >=Stealth] (trwb) -- (w1i);
  \draw[black!50, dotted, thick, ->, >=Stealth] (trwb) -- (wi);

  \draw[blue!50, dotted, thick, ->, >=Stealth] (w1) -- ($(w1i)+(0, \arrowendy)$);
  \draw[blue!50, dotted, thick, ->, >=Stealth] (w) -- ($(wi)+(0, \arrowendy)$);

  \draw[blue, thick, ->, >=Stealth] (W1) -- node[above] {$\succeq$} (w1i);

  \draw[blue, thick, ->, >=Stealth] (W) -- node[left] {$\preceq$} (wi);

\end{tikzpicture}
\caption{An illustration of our setting (in black), the problem of spurious splittability (in red), and the backward satisfiability solution (in blue).
The black dashed arrows represent the state encoding function $\trs{\cdot}$ from front-end states to IVL states, and the black dashed ellipse represents the image of $\trs{\cdot}$.
The red arrows represent a possible spurious split of the IVL state $\trs{\omega}$ into $\Omega_A$ (that satisfies the translated assertion $\tra{A}$) and $\Omega_F$ (the frame), that have no front-end counterparts.
The blue arrows show how the backward satisfiability of $A$ guarantees the existence of front-end states $\omega_A$ and $\omega_F$,
such that $\omega_A$ satisfies $A$, $\trs{\omega_A}$ is smaller or equal to $\Omega_A$, and $\trs{\omega_F}$ is larger or equal to $\Omega_F$.}

\label{fig:bsat}
\end{figure*}

\subsection{State Encoding Relations}
\label{subsec:relation}
\label{subsec:homomorphic-enc}
                                   
Discrepancies between front-end and IVL states can stem from many different sources:
a different heap structure (\eg in \figref{fig:dfc-ex}),
a different permission model (\eg binary vs.\ fractional permissions), or even a different representation of the same permission model (\eg rational vs.\ real numbers for fractional permissions).
To abstract over these different aspects
and make our approach generically applicable to different front-ends and IVLs,
our framework is parametric in the front-end and IVL state models,
and we simply assume the existence of a state encoding relation $\rels{\cdot}{\cdot}$ between them.
Intuitively, $\rels{\omega}{\Omega}$
holds when the IVL state $\Omega$ represents the front-end state $\omega$.

\paragraph{Relations from encoding functions}
The state encoding \emph{relation} is often derived naturally from a \emph{function} $\trs{\cdot}$ that encodes a front-end separation logic state into the corresponding IVL state, in which case $\rels{\omega}{\Omega}$ is defined as $\Omega = \trs{\omega}$.
For example, $\trs{\cdot}$ could encode binary permissions into fractional permissions by setting a permission of $1$ for locations owned by the state, and $0$ for the others.
In the example from \figref{fig:dfc-ex}, our function $\trs{\cdot}$ encodes states as follows:
The ownership of a non-existent field \fecode{f} of a created object \fecode{x} is encoded as the ownership of $\addedf{}$ of \vcode{x} with its value set to \vfalse; the ownership of a created field \fecode{f} with value $v$ is encoded as the ownership of both $\addedf{}$ and $\ivlf{}$, where the former has value \vtrue{} and the latter stores the actual value $v$.
The situation is represented in black in \figref{fig:bsat}:
Our function $\trs{\cdot}$ encodes the front-end states into (typically) a subset of the IVL states.
For example, the image of the encoding function from binary permissions is the subset of IVL states with permissions $0$ or $1$.

\paragraph{The need for a relation}
State encoding functions are too restrictive to capture some front-end encodings, for instance, translations that encode duplicable permissions into fractional permissions,
as no fixed positive fraction, however small, can be duplicated.
However, duplicable permissions can be interpreted in the IVL as \emph{any arbitrary} positive fraction: A front-end state with a single duplicable resource then corresponds
to infinitely many IVL states, each with a different fraction.
Such encodings
are easily captured by our binary relation $\rels{\cdot}{\cdot}$, as we will show in \secref{sec:instantiations}.

\subsection{The Homomorphism Requirement}
\label{subsec:homomorphism}

At the core of separation logic is the concept of \emph{separation}:
A state $\omega$ satisfies the assertion $P * Q$ if it can be split into two states $\omega_P$ and $\omega_Q$ such that $\omega_P$ satisfies $P$, $\omega_Q$ satisfies $Q$, and $\omega = \omega_P \oplus \omega_Q$, where $\oplus$ is a partial commutative and associative binary operation.
In particular, both front-end and IVL states are equipped with their operator $\oplus$.
For the translation to be meaningful, we require the state encoding relation to be homomorphic with respect to $\oplus$.

For simplicity, we describe here the homomorphism requirement for the case where $\rels{\cdot}{\cdot}$ is defined by an encoding function $\trs{\cdot}$, but we adapt it later to the general case of a relation.
The encoding function $\trs{\cdot}$ is required to be homomorphic:
For all front-end states $\omega_1$ and $\omega_2$,
$\trs{\omega_1 \oplus \omega_2} = \trs{\omega_1} \oplus \trs{\omega_2}$.
This is necessary so that the separating conjunction ($*$) in the IVL program (\eg when adding or removing resources via inhale or exhale) corresponds to the separating conjunction in the front-end program
(\eg in its parallel rule).



\paragraph{Unsoundness without homomorphism}
To see why the homomorphism requirement is necessary, consider encoding a front-end state model with counting permissions into an IVL state model with fractional permissions.
Counting permissions~\cite{bornatPermissionAccountingSeparation2005} is a permission model that allows one to subtract from a full permission an unbounded number of units.
In this model, a state holds either $1-k\cdot\epsilon$ or $k\cdot\epsilon$ permission to a heap location (for any non-negative $k$), where a full permission $1$ is required to write to a location, and any positive number of units $k\cdot\epsilon$ provides read access.
Counting permissions are, for instance, used in the Chalice verifier~\cite{leinoBasisVerifyingMultithreaded2009}.

A simple (but unsound) encoding of counting permissions into an IVL offering fractional permissions~\cite{boylandCheckingInterferenceFractional2003}, such as VeriFast and Viper,
is to encode $k \cdot \epsilon$ as the fraction $1 - \frac{1}{2^k}$, and $1 - k \cdot \epsilon$ as the complementary fraction $\frac{1}{2^k}$.
This encoding ensures that the sum of the fractions for $1-k\cdot\epsilon$ and $k\cdot\epsilon$ provides write permission and, for positive $k$, each fraction is positive and, thus, provides read permission.
However, this encoding is \emph{not} homomorphic:
Let $\omega^k$ be a front-end state with $k \cdot \epsilon$ permission to some heap location \code{x.f},
and $\Omega^p$ be an IVL state with a fractional permission $p$ to the corresponding heap location.
Then $\trs{\omega^1 \oplus \omega^1} = \trs{\omega^2} = \Omega^{1 - \frac{1}{2^2}} = \Omega^{\frac{3}{4}}$, whereas
$\trs{\omega^1} \oplus \trs{\omega^1} = \Omega^{1 - \frac{1}{2^1}} \oplus \Omega^{1 - \frac{1}{2}} = \Omega^1 \neq \Omega^{\frac{3}{4}}$.
In other words, the encoding of the sum of the front-end states $\omega^1 \oplus \omega^1$
has \emph{strictly less} permission than the sum of the individual encodings.

\begin{figure}
\footnotesize
\begin{minipage}{0.5\textwidth}
\begin{frontend}[mathescape]
method m(x: Ref, y: Ref, z: Ref)
  requires x.f $\stackrel{\epsilon}{\mapsto}$ _ * y.f $\stackrel{\epsilon}{\mapsto}$ _ * z.f $\stackrel{\epsilon}{\mapsto}$ _
{
  if (x = y $\wedge$ x = z)
    assert false
}
\end{frontend}
\end{minipage}
\begin{minipage}{0.5\textwidth}
\begin{viper}[mathescape]
method m(x: Ref, y: Ref, z: Ref) {
  inhale x.f $\stackrel{\frac{1}{2}}{\mapsto}$ _ * y.f $\stackrel{\frac{1}{2}}{\mapsto}$ _ * z.f $\stackrel{\frac{1}{2}}{\mapsto}$ _
  if (x = y $\wedge$ x = z)
    assert false
}
\end{viper}
\end{minipage}
    \caption{Encoding counting permissions $k \cdot \epsilon$ as fractions $1 - \frac{1}{2^k}$ is unsound. The front-end program on the left permits the three inputs to be aliased, but the IVL program on the right does not.}
    \label{fig:rd-unsound-1}
\end{figure}

This violation of the homomorphic requirement results in an unsound front-end translation, as illustrated by the example in \figref{fig:rd-unsound-1}.
The front-end program on the left might lead to an assertion violation because the counting permissions required in the precondition do \emph{not} prevent all three inputs from being aliased, as there is no bound on the number of units for each location.
In contrast, the assertion inhaled by the IVL program (on the right) rules out the case that all three inputs are aliased since it is not possible to own more than one full permission to each location.
Therefore, the IVL program verifies even though the input program should not. 

Note that this unsoundness is independent of the particular fraction chosen to represent one unit, as long as this fraction is strictly positive;
smaller fractions just need more aliases to exhibit the unsoundness.
\citet{SummersM20} use an alternative encoding, in which a unit is encoded as a \emph{symbolic} fraction, together with the axioms $f > 0$ and $\forall k \in \mathbb N.\, k \cdot f < 1$. This representation makes it harder for an SMT-based verifier to \emph{exploit} the unsoundness, but it is still present as there is no fraction $f$ that satisfies these axioms.

\subsection{Spurious Splittability}
\label{subsec:spurious-splittability}

As discussed in \secref{sec:intro}, front-end translations rely on SL operations to add resources to the verification state (\emph{produce}) and to remove them (\emph{consume}). IVLs provide these operations (\eg \vcode{inhale} and \vcode{exhale} in Viper) to encode SL proof rules. For example, the front-end translation from \figref{fig:dfc-ex} uses these operations
to model allocation and parallel composition, which are not supported by the IVL\@.

In general, a front-end statement $C$ (\eg \fecode{x := alloc(f)}) can be modeled in the IVL via the corresponding rule from the front-end separation logic:
If the triple $\{P\} \; C \; \{Q\}$ is valid in the front-end separation logic, then a possible (and typical) translation for $C$ in the IVL is
\vcode{exhale P; inhale Q} (assuming that $C$ does not modify local variables for simplicity).
Starting in the IVL state $\Omega$, \vcode{exhale P} will split $\Omega$ into $\Omega_P \oplus \Omega_F$ (where $\Omega_P$ satisfies $P$) and remove $\Omega_P$. The resulting state $\Omega_F$ represents the frame, \ie the resources that are unaffected by the execution of $C$\@.
The \vcode{inhale Q} statement will then add another state $\Omega_Q$ to the frame $\Omega_F$, resulting in the state $\Omega_Q \oplus \Omega_F$.
The state $\Omega_Q$, which satisfies $Q$, models how the statement $C$ modified $\Omega_P$.
Intuitively, this encoding is justified by the frame rule of the front-end separation logic, with $\Omega_F$ used as frame.

Unfortunately, this intuitive justification might not be valid, because the split $\Omega_P \oplus \Omega_F$ might be \emph{spurious}:
As represented in \figref{fig:bsat} (in red),
the IVL states $\Omega_P$ and $\Omega_F$ might not have front-end counterparts,
even if the initial IVL state $\Omega$ has a front-end counterpart (\ie a front-end state $\omega$ such that $\Omega = \trs{\omega}$).
In such a case, the frame rule of the front-end separation logic cannot be used soundly with the IVL state $\Omega_F$, as it has only been proven sound for front-end states.
Such spurious splits arise because the front-end states generally correspond to a \emph{subset} of the IVL states;
in other words, there are generally infinitely many more IVL states than front-end states,
and thus infinitely many IVL states without front-end counterparts.
We call IVL states that have a front-end counterpart \emph{concrete} states, and \emph{spurious} those that do not.

\begin{figure}
\footnotesize
\begin{minipage}{0.5\textwidth}
\begin{frontend}[mathescape]
method main() {
  x := alloc(f)
  fork reader(x)
  y := x.f
}

method reader(x: Ref)
  requires $\code{x.f} \mapsto \_$
{
  var v := x.f
  ...
}
\end{frontend}
\end{minipage}
\begin{minipage}{0.5\textwidth}
\begin{viper}[mathescape]
method main() {
  havoc x
  inhale $\code{x.f} \mapsto \_$
  exhale $\exists p > 0 \ldotp \code{x.f} \stackrel{p}{\mapsto} \_$
  y := x.f
}

method reader(x: Ref) {
  inhale $\exists p > 0 \ldotp \code{x.f} \stackrel{p}{\mapsto} \_$
  var v := x.f
  ...
}
\end{viper}
\end{minipage}
\caption{Encoding binary permissions using fractional permissions is generally unsound.
The IVL program on the right allows reading \code{x.f} after the fork because of the use of fractional permissions, and thus is correct.
In contrast, the front-end program uses binary permissions, and thus should not be verified.}
\label{fig:concurrent-reads-unsound}
\end{figure}

\paragraph{Unsoundness from spurious splits}
As a simplified example (we show more examples below),
consider a front-end that uses binary permissions. The encoding into an IVL with fractional permissions  attempts to improve framing by determining with a simple pre-analysis whether a method updates a heap location. If not, it encodes permission to that location as an arbitrary positive fraction rather than full permission, such that callers can frame the value around calls. In this setting, the front-end program on the left of \figref{fig:concurrent-reads-unsound} does not verify because there is no permission to access \fecode{x.f} after the fork. However, the encoding on the right verifies because the assertion $\exists p > 0 \ldotp \code{x.f} \stackrel{p}{\mapsto} \_$ in the IVL program (on the right of \figref{fig:concurrent-reads-unsound}) can be satisfied by any positive fraction, enabling the read \vcode{y := x.f} in the main thread. This unsoundness is caused by spurious splittability: an IVL state with full permission to \fecode{x.f} can be split into two states, each with $1/2$-permission (one being passed to the forked thread and the other retained to read \fecode{x.f} subsequently). However, these states have no corresponding front-end states, such that the same reasoning steps are not valid there.
We show in \secref{subsec:unsoundness} that similar problems arise even with an assertion syntax suitably restricted for the front-end state model.

\subsection{Backward Satisfiability}
\label{subsec:backward-satisfiability}

The unsoundness in the previous example comes from the fact that the
encoded precondition $\exists p > 0 \ldotp \code{x.f} \stackrel{p}{\mapsto} \_$ can be satisfied by a spurious IVL state, which results in a spurious frame. In particular, no smaller concrete state satisfies the encoded precondition, such that \emph{any} split
that enables verifying the whole IVL program is spurious.

We prevent this source of unsoundness by requiring front-end assertions to satisfy a novel soundness condition.
Intuitively, a front-end assertion $A$ is \emph{backward satisfiable} if for any%
\footnote{More precisely, backward satisfiability only considers IVL states $\Omega_A$ that are smaller than the encoding $\trs{\omega}$ of some front-end state $\omega$, \ie such that $\trs{\omega} = \Omega_A \oplus \Omega_R$ for some IVL state $\Omega_R$.}
IVL state $\Omega_A$ that satisfies its translation $\tra{A}$,
there exists a front-end state $\omega_A$ that satisfies $A$ and whose encoding $\trs{\omega_A}$ is smaller than $\Omega_A$.
Intuitively,
and as depicted in \figref{fig:bsat} in blue,
backward satisfiability guarantees the existence of an alternative, non-spurious split $\trs{\omega} = \trs{\omega_A} \oplus \trs{\omega_F}$ for some front-end states $\omega_A$ (satisfying $A$) and $\omega_F$, such that $\trs{\omega_F} \succeq \Omega_F$.
The latter ensures that the IVL frame $\Omega_F$ can be used soundly with the frame rule from the front-end SL.%
\footnote{Throughout the paper, we assume an affine front-end SL.}

Importantly, backward satisfiability holds for many basic SL assertions,
such as points-to $\code{x.f} \stackrel{p}{\mapsto} v$ and pure assertions (\ie without permissions).
Backward satisfiability is preserved by many important SL connectives.
For example, if $A$ and $B$ are backward satisfiable, then so is $A * B$ (with the straightforward definition $\tra{A * B} \triangleq \tra{A} * \tra{B}$).
Unfortunately, backward satisfiability is \emph{not} preserved by all connectives, such as magic wands (separating implications), as we show next.

\subsection{Unsoundness without Backward Satisfiability}
\label{subsec:unsoundness}

In this subsection, we show two examples that
violate backward satisfiability because of the magic wand connective, and how the soundness of front-end translations can be compromised when the backward satisfiability condition is violated.
The framework presented in this paper allows one to uncover such subtle sources of unsoundness and to prove that correct encodings are indeed sound.

\paragraph{Encoding binary permissions into fractional permissions.}
The example in \figref{fig:binary-frac-unsound} encodes the input program on the left into the identical IVL program on the right.
The front-end program on the left uses binary permissions, whereas the IVL encoding on the right uses fractional permissions.
The assertion $W$ is defined using a magic wand (separating implication) connective:

    {
\vspace{-3.5mm}
      \small
    \begin{align*}
      W &\triangleq \febasic{x.f} \mapsto \_ \wand (\febasic{x.f} \mapsto \_ * \febasic{y.g} \mapsto \_)
    \end{align*}
    }
Intuitively, it expresses that one can give up the wand as well as ownership of \vcode{x.f} to obtain ownership of both \vcode{x.f} and \vcode{y.g} (that is, apply modus ponens). 

\begin{figure}[ht]
\footnotesize
\begin{minipage}{0.5\textwidth}
\begin{frontend}[mathescape]
method m(x: Ref, y: Ref)
  requires x.f $\mapsto$ _
  ensures $W*W*W$
{ $\feskip$ }
$\phantom{ }$
\end{frontend}
\end{minipage}
\begin{minipage}{0.5\textwidth}
\begin{viper}[mathescape]
method m(x: Ref, y: Ref) {
  inhale x.f $\mapsto$ _
  $\vskipp$
  exhale $W*W*W$
}
\end{viper}
\end{minipage}
\caption{Encoding binary permissions using fractional permissions is generally unsound. The IVL program on the right verifies because there are fractional permissions for the footprints of the three magic wands $W$. With the binary permissions used by the front-end program on the left, it is not possible to provide sufficient permissions for all three wands.}
\label{fig:binary-frac-unsound}
\end{figure}

Magic wands themselves represent resources (the wand's \emph{footprint}) that, combined with the left-hand side of the wand, entail the right-hand side. With binary permissions, $W$'s footprint contains either permission to \fecode{x.f} (so that combining it with the left-hand side yields false, thereby trivially satisfying the right-hand side) or permission to \fecode{y.g}. No matter which footprint is chosen, the postcondition represents more than full permission to a heap location and is thus unsatisfiable. Hence, verification should fail.
Nevertheless, verification of the IVL program succeeds. With fractional permissions, one can choose $W$'s footprint as
$\febasic{x.f}\stackrel{1/3}{\mapsto}\febasic{\_}$, such that the right-hand side is entailed trivially. With this choice, the postcondition holds, and verification succeeds. 

This unsoundness stems from the violation of backward satisfiability by $W$ (despite the fact that both sides of the magic wand are backward satisfiable): the IVL may satisfy the translation of $W$ in a state $\Omega_A$ with $\frac{1}{3}$ permission to \vcode{x.f}. However, the \emph{only} front-end state $\omega_A$ that satisfies $W$ has full permission to \vcode{x.f} and, thus, its encoding is \emph{not} smaller than or equal to $\Omega_A$.

This example also demonstrates that a commonly-used state encoding is sound only if it is combined with appropriate restrictions on the allowed assertions. For instance, disallowing binary permissions inside magic wands would rule out our unsound example.

\paragraph{Encoding rational permission amounts into real permission amounts.}
Our next example shows that the common practice of encoding SLs with rational permission amounts into automatic verifiers targeting real permission amounts is surprisingly unsound in the presence of the magic wand connective.
Consider the following two assertions:

    {
\vspace{-1.5mm}
      \small
\[
        \begin{aligned}
        P_1 &\triangleq \left(\existsntp{v} \febasic{y.g} \mapsto v * 0 < v * v^2\leq \frac{1}{2}\right) \wand \left(\existsntp{v} \febasic{y.g} \mapsto v * \febasic{x.f} \stackrel{v}{\mapsto} \_\right)\\
        P_2 &\triangleq \left(\existsntp{v} \febasic{y.g} \mapsto v * 0 < v < 1 * (1-v)^2 \geq \frac{1}{2}\right) \wand \left(\existsntp{v} \febasic{y.g} \mapsto v * \febasic{x.f} \stackrel{v}{\mapsto} \_\right)
        \end{aligned}
\]
    }

In both of them, the left-hand side of the magic wand constrains the value $v$, which is used as a permission amount for \code{x.f} on the right-hand side. The predicate $\febasic{y.g} \mapsto v$ enforces the values $v$ on both sides to be the same. $P_1$ constrains $v$ to be in the half-open interval $(0,\frac{1}{\sqrt 2}]$, whereas $P_2$ allows values for $v$ in $(0,1-\frac{1}{\sqrt 2}]$. In both wands, the left-hand side provides no permission for \code{x.f}, so in both cases, $\febasic{x.f} \stackrel{v}{\mapsto} \_$ is part of the footprint of the wand.

\begin{figure}[ht]
\footnotesize
\begin{minipage}{0.5\textwidth}
\begin{frontend}[mathescape]
method m(x: Ref, y: Ref)
  requires x.f $\mapsto$ _
  ensures $P_1*P_2$
{ $\feskip$ }
$\phantom{ }$
\end{frontend}
\end{minipage}
\begin{minipage}{0.5\textwidth}
\begin{viper}[mathescape]
method m(x: Ref, y: Ref) {
  inhale x.f $\mapsto$ _
  $\vskipp$
  exhale $P_1*P_2$
}
\end{viper}
\end{minipage}
\caption{Encoding rational permission amounts as real numbers is unsound. The front-end program on the left and the IVL program on the right interpret the permission amounts in the postcondition differently, such that the IVL program verifies successfully even though the front-end program should not.}
\label{fig:rational-real-wand-unsound}
\end{figure}

The front-end and IVL methods in \figref{fig:rational-real-wand-unsound} require full ownership to \code{x.f} and ensure the separating conjunction $P_1*P_2$. The two methods look identical, but use rationals and reals as permission amounts, respectively. 
The footprints of both magic wands need to contain $\febasic{x.f} \stackrel{v}{\mapsto} \_$ for \emph{every} value $v$ permitted by the left-hand side. The IVL program, which uses real numbers as permission amounts, thus, uses $\frac{1}{\sqrt 2}$ for $P_1$ and $1-\frac{1}{\sqrt 2}$ for $P_2$. In real arithmetic, these add up to a full permission, which is provided by the precondition. Hence, verification of the IVL program succeeds. 
However, the front-end logic, which uses rational permission amounts, cannot represent the amounts $\frac{1}{\sqrt 2}$ and $1-\frac{1}{\sqrt 2}$ precisely and, thus, needs to conservatively over-approximate them. As a result, their sum exceeds the full permission provided in the precondition, which causes verification to fail. 

The unsoundness in this example again stems from the violation of backward satisfiability, by both $P_1$ and $P_2$ (despite the fact that all sides of the magic wands are backward satisfiable): the IVL may satisfy $P_1$ (resp. $P_2$) in a state $\Omega_A$ with $\frac{1}{\sqrt{2}}$ (resp. $1-\frac{1}{\sqrt{2}}$) permission to \vcode{x.f}, while the front-end must provide strictly more amount of permission for $\omega_A$ to satisfy $P_1$ (resp. $P_2$).

This example shows that encoding rational permission amounts as reals is generally unsound.
Since it should in principle be possible to formalize fractional permissions with real numbers instead of rational ones, we do not expect this unsoundness to compromise actual verification results.
However, the mismatch makes it impossible to prove a formal soundness result for front-ends using this common state model encoding.

\revision{These unsoundnesses stem from the mismatch between the front-end and IVL state models: when the two coincide, the state encoding can be taken to be the identity, backward satisfiability holds trivially, and \emph{all} connectives (including magic wands and universal quantifications) can be translated soundly. Conversely, for front-ends with a sufficiently expressive assertion language, backward satisfiability is necessary: whenever an assertion's translation violates it \emph{and} the front-end can express the corresponding upper bound and frame (as in all our examples), one can construct an unsound method as above, in which the IVL translation verifies but the front-end program does not.}

\section{A Framework for Proving Front-End Translations Sound}
\label{sec:framework}

Our core contribution is a formal framework that allows one to prove that the state model encoding of a front-end translation is sound.
In this section, we formalize our framework: the semantic model of the IVL (\secref{subsec:preliminaries}), its parameters (\secref{subsec:syntax}) and their soundness requirements (\secref{subsec:assumption}), our soundness result (\secref{subsec:result}), and a generalization in which the state encoding is a relation rather than a function (\secref{subsec:generalize-fw}). All results are formalized in \isabelle{}\@. Although the formalization uses Viper as the IVL, all requirements and results are independent of any specific IVL, so we present them for a general SL-based IVL.

\subsection{Preliminaries}
\label{subsec:preliminaries}

We adopt much of our semantic model of the IVL from \citet{dardinierFormalFoundationsTranslational2025}.

\paragraph{IDF algebras.}
To make our framework widely applicable, our formalization captures both IVLs based on standard SL (\eg VeriFast) and IVLs based on implicit dynamic frames (IDF) \cite{smansImplicitDynamicFrames2012,SpiesMZSLMD25} (\eg Viper).
To achieve that, we formalize states as IDF algebras, a generalization of separation algebras that also captures IDF state models:%
\footnote{\revision{Compared with the definition from \citet{dardinierFormalFoundationsTranslational2025}, we make the $\unitNa$ function explicit and independent from $\stabilizeNa$ and $\core{\cdot}$; the definition from \citet{dardinierFormalFoundationsTranslational2025} can be derived from ours by setting $\unit{s} \triangleq \stabilize{\core{s}}$. Our generalization is crucial to capture front-end state models where $\stabilize{\core{\cdot}}$ does not satisfy the properties for $\unit{\cdot}$, \eg the immutable heaps from \secref{subsec:immutable-heap-inst}, where $\stabilize{\core{h}}=h$ but $\unit{h}=\lambda l.\, \bot$.}}

\begin{definition}[IDF algebra]\label{def:idf-algebra}
    An IDF algebra is a tuple $(\Sigma,\oplus, \core{\cdot}, \stableNa, \stabilizeNa, \unitNa)$ that satisfies a set of axioms (listed in \appendixref{appsec:framework}{A}), where $\Sigma$ is a set of states, $\oplus:\Sigma \times \Sigma \rightharpoonup \Sigma$ is a partial, commutative, and associative addition on $\Sigma$, $\core{\cdot}$, $\stabilizeNa$, and $\unitNa$ are endomorphisms of $\Sigma$, and $\stableNa$ is a predicate on $\Sigma$.
\end{definition}

Intuitively, $\oplus$ denotes the composition of states, $\core{\cdot}$ projects a state on its largest duplicable part,
$\stableNa$ decides the self-containedness of a state (\ie the state holds enough ownership to all its fields so that none of them may be modified by the environment),
$\stabilizeNa$ projects a state on its largest stable part, and $\unitNa$ denotes the unit element of $\oplus$.

IDF assertions separate ownership information from value constraints. For instance, SL's points-to assertion $\febasic{x.f} \mapsto \febasic{v}$ is expressed in IDF as $\febasic{acc(x.f)} * \febasic{x.f = v}$. To ensure sound framing, IDF requires assertions to be self-framing; that is, an assertion may constrain a heap location only if it includes the permission to that location. Consequently, 
$\febasic{acc(x.f)} * \febasic{x.f = v}$ is self-framing, but
$\febasic{x.f = v}$ alone is not. We defined self-framedness of assertions and other related concepts as follows:

\begin{definition}\label{def:self-framedness}
Let $P$ be an IDF assertion (\ie a set of states from an IDF algebra).
\begin{itemize}
    \item $P$ is \emph{self-framing}, written $\selfFraming{P}$, iff $\forallntp{ \omega} \omega \in P \Leftrightarrow \stabilize{\omega} \in P$.
    \item A state $\omega$ \emph{frames} $P$, written $\framedBy{\omega}{P}$, iff $\selfFraming{\{\omega\} * P}$.
    \item An IDF assertion $B$ \emph{frames} $P$, written $\framedBy{B}{P}$, iff $\forallntp{\omega \in B} \stable{\omega} \Rightarrow \framedBy{\omega}{P}$.
    \item \emph{$P$ frames an expression} (\ie a partial function from states to values) $e$, written $\framedBy{P}{e}$, iff $e(\omega)$ is defined for all $\omega \in P$.
    \item \emph{An expression $e$ is stable under $P$}, written $\stableUnder{e}{P}$, iff $\forall \omega \in P$, $e(\omega)$ is defined iff $e(\stabilize{\omega})$ is defined, and $e(\omega)=e(\stabilize{\omega})$ when they are both defined.
\end{itemize}
\end{definition}

\paragraph{IVL semantics.} An IVL state consists of a store of local variables and a customized heap $\varphi \in \Sigma$ equipped with an IDF algebra structure, \ie the set of IVL states $\Sigma_V=(\Variable \rightharpoonup V) \times \Sigma$, where $V$ is the set of values and the store $\Variable \rightharpoonup V$ is instantiated to the agreement algebra, \ie addition on stores is defined only for identical stores and yields the argument store, $\core{\cdot}$, $\stabilizeNa$, and $\unitNa$ are instantiated to the identity function, and $\stable{s}$ holds for any store $s$.

We require that the IVL supports three primitive statements $\vhavoc{x}$, $\vinhale{A}$, $\vexhale{A}$, and sequential compositions $C_1;C_2$ and conditional statements $\vif{b}{C_1}{C_2}$ in order to translate some front-end control structures. Intuitively, $\vhavoc{x}$ erases the value of local variable $x$ by overwriting it non-deterministically, $\vinhale{A}$ adds the resource specified by assertion $A$ into the current state and assumes the value constraints in $A$, and $\vexhale{A}$ checks that $A$ holds in the current state and subsequently removes the resource corresponding to $A$ from the current state.

We assume an axiomatic semantics for the IVL. The axiomatic semantic judgements are of the form $\axiomSem{P}{C}{Q}$, where $\Delta$ is a type context\footnote{
We omit typing details here, but our \isabelle{} formalization includes them: it requires the state encoding to preserve well-typedness and deals only with well-typed states (whose store and heap contain values of the types prescribed by $\Delta$). All states in \secref{sec:framework} and \secref{sec:instantiations} are implicitly assumed well-typed.}, $P$ and $Q$ are IVL semantic assertions (\ie sets of IVL states), and $C$ is an IVL statement. The axiomatic semantic rules of the aforementioned statements are given in \appendixref{appsec:framework}{A}.

\subsection{Framework Parameters}
\label{subsec:syntax}

To capture a wide range of front-ends, our framework offers numerous parameters that describe the input language to the front-end as well as the translation performed by the front-end.

\paragraph{Parameters describing the input language}
The framework parameters for the input language describe, for instance, the available boolean expressions, assertions, and statements. \figref{fig:fe-input} provides an overview of the parameters; in the following, we explain some of the most interesting ones; further details are included in the appendix.

\begin{figure}
    \footnotesize
    \begin{tabularx}{\textwidth}{|c|X|}
        \hline
        \textbf{Framework parameter} & \multicolumn{1}{c|}{\textbf{Description}}\\
        \hline
        $T_F$ & Set of front-end types.\\
        \hline
        $V_F$ & Set of front-end values.\\
        \hline
        $\tp:V_F \to T_F$ & Function that assigns each value $v \in V_F$ a type $t \in T_F$.\\
        \hline
        $H_F$ & Set of front-end \emph{ghost heaps} used for reasoning about the resources in front-end states. They must be instantiated as an IDF algebra.\\
        \hline
        $H_P$ & Set of front-end \emph{program heaps}.\\
        \hline
        $\embedh:H_P \to H_F$ & Function that embeds program heaps onto ghost heaps.\\
        \hline
        $BE_F$ & Set of front-end boolean expressions.\\
        \hline
        $\interpbena:BE_F \to (\Sigma_F \rightharpoonup \{\fekeyword{true},\fekeyword{false}\})$ & Interpretation for each boolean expression $b \in BE_F$ as a partial predicate on front-end ghost states.\\
        \hline
        $\fvbe:BE_F \to \mathcal{P}(\Variable)$ & (Possibly overapproximating) finite free variable set of a boolean expression.\\
        \hline
        $A_{Fp}$ & Set of input assertions.\\
        \hline
        \revision{$\wfa^{\Delta_F}(\cdot)$} & \revision{Well-formedness predicate on input assertions in a given type context $\Delta_F$.}\\
        \hline
        $\interpana:A_{Fp} \to \mathcal{P}(\Sigma_F)$ & Interpretation of each input assertion $A_p \in A_{Fp}$ as a set of front-end ghost states satisfying the assertion under type context $\Delta_F$.\\
        \hline
        $\fva:A_{Fp} \to \mathcal{P}(\Variable)$ & (Possibly overapproximating) finite free variable set of an input assertion.\\
        \hline
        $C_{Fp}$ & Set of primitive statements.\\
        \hline
        $\wfc^{\Delta_F}(\cdot)$ & Well-formedness predicate on primitive statements in a given type context $\Delta_F$.\\
        \hline
        Rules of the form $\sigma \stackrel{c}{\to}_{\Delta_F} \sigma'$ & Operational semantic reduction rules for each primitive statement $c \in C_{Fp}$. $\sigma,\sigma' \in \Sigma_P$ are program pre- and post-states, and $\Delta_F$ is a type context.\\
        \hline
        Rules of the form $\cslp{\Delta_F}{P}{c}{Q}$ & CSL rules for primitive statements $c \in C_{Fp}$. $\Delta_F$ is a front-end type context, and $P$ and $Q$ are front-end semantic assertions (\ie sets of front-end ghost states).\\
        \hline
        $\wrc:C_{Fp} \to \mathcal{P}(\Variable)$ & (Possibly overapproximating) finite modified variable set of a primitive statement.\\
        \hline
        $\fvc:C_{Fp} \to \mathcal{P}(\Variable)$ &  (Possibly overapproximating) finite free variable set of a primitive statement.\\
        \hline
    \end{tabularx}
    \caption{Framework parameters for the input language.}
    \label{fig:fe-input}
\end{figure}

Types and boolean expressions are fully described by the respective framework parameters ($T_F$ and $BE_F$). The parameter $A_{Fp}$ for assertions describes a set of so-called \emph{input assertions}; the assertions supported by the front-end are then obtained using the following grammar, where $A_p \in A_{Fp}$, $b \in BE_F$, $t \in T_F$\revision{, and $P$ is a designated symbol for recursive occurrences (as we explain below):}
{
    \small
  \[
  A \Coloneqq \;
  A_p \mid b \mid A*A \mid A \vee A \mid b \Rightarrow A \mid \existstp{x}{t} A \revision{ \mid P \mid \lfp{A}}
  \]
}%
That is, separating conjunctions, disjunctions, implications, existential quantifiers\revision{, and inductive predicates}
are built into the framework,
\revision{because}
they preserve backward satisfiability (as we present in \lemref{thm:bsat-asst}) and we can thus reduce the effort of users of our framework to only proving backward satisfiability for input assertions; all other SL connectives (\eg magic wands) have to be provided as input assertions.
\revision{Following \citet{dardinierFractionalResourcesUnbounded2022},
we formalize inductive predicates with one designated symbol $P$, which represents a recursive occurrence inside the least fixed-point computation $\lfp{A}$.
For example, the standard linked-list inductive predicate can be expressed as}%
\footnote{\revision{The standard linked-list predicate is usually defined as
$\textit{List(x)} \triangleq \febasic{x} \ne \fenull \Rightarrow (\febasic{acc(x.val)} * \febasic{acc(x.next)} * \textit{List}(\febasic{x.next}))$.
In our syntax, the recursive occurrence $P$ (which in our example represents $\textit{List}$) does not take explicit arguments as input;
we thus use an existential quantifier to bind the heap-dependent expression $\febasic{x.next}$ to the implicit argument $\febasic{x}$.}}
{
    \small
$$\lfp{\febasic{x} \ne \fenull \Rightarrow (\febasic{acc(x.val)} * \febasic{acc(x.next)} * (\existstp{v}{\fekeyword{Ref}} \febasic{v = x.next} * (\existstp{x}{\fekeyword{Ref}} \febasic{x = v} * P)))}$$
}
$\interpana$ is extended to an interpretation function on all front-end assertions using the standard interpretation for the built-in connectives \revision{and the least fixed point interpretation for inductive predicates}. $\fva$, the function that returns the set of free variables for assertions,
\revision{and $\wfa^{\Delta_F}$, the well-formedness predicates of assertions,}
are extended to all front-end composite assertions accordingly. The concrete definitions of the extensions of $\interpana$, $\fva$
\revision{, and $\wfa^{\Delta_F}$}
are in \appendixref{appsec:framework}{A}.

Similarly, the parameter for statements $C_p$ provides a set of \emph{primitive} statements, from which the framework derives the set of statements using the following grammar, where $C_p \in C_{Fp}$, $b \in BE_F$, and $A$ is a front-end assertion:
{
    \small
    \[
    C \Coloneqq \;
    \feskip \mid C_p \mid C; C
    \mid \feif{b}{C}{C}
    \mid \fewhile{b}{A}{C}
    \mid \fepar{A}{C}{A}{A}{C}{A}
    \]
}
Loops and parallel compositions contain assertions (invariants for loops, and pre- and postconditions for each parallel branch) for verification purposes. $\wfc$, $\fvc$, and $\wrc$ can be extended to all front-end statements in a straightforward way, as shown in \appendixref{appsec:framework}{A}.

The small-step operational semantic reduction rules and CSL rules for these control structures are fixed to the standard ones in the framework, as shown in \appendixref{appsec:framework}{A}. For primitive statements, they are expressed by the framework parameters.

To customize the state model of the input language, the framework has parameters for types ($T_F$), values ($V_F$), program heaps ($H_P$), and ghost heaps ($H_F$). Intuitively, program heaps abstract the memory part of execution states  and are involved in the definition of the operational semantics of the front-end language. In contrast, ghost heaps typically include the program heaps and additional SL resources (\eg permission masks) for reasoning under SL\@. They are thus elements of an IDF algebra. We define front-end ghost states $\Sigma_F=(\Variable \rightharpoonup V_F) \times H_F$ and program states $\Sigma_P=(\Variable \rightharpoonup V_F) \times H_P$ to consist of a store of local variables and a ghost (resp.\ program) heap. The embedding function $\embedh$ of heaps can then be naturally extended to an embedding $\embeds:\Sigma_P \to \Sigma_F$ of program states onto ghost states.

\paragraph{Parameters describing the front-end translation}
Our framework captures the translation from the front-end language to the IVL by parameters $\trt{\cdot}$, $\trv{\cdot}$, $\trh{\cdot}$, $\trctxt{\cdot}$, $\trbe{\cdot}$, $\trap{\cdot}$, and $\trcp{\cdot}$ for the translation of front-end types, values, ghost heaps, type contexts, boolean expressions, input assertions, and primitive statements, respectively. The functions $\trap{\cdot}$ and $\trcp{\cdot}$ may translate front-end input assertions and primitive statements into composite ones in the IVL\@.

The translation of types and values may in general lose information. For example, when translating the front-end type \fekeyword{Rat} of rational values into the IVL type \vkeyword{Real} of real values, the rationality of front-end values will not be reflected in their IVL translation. As a result, witnesses for existential quantifiers in the IVL do not necessarily witness the corresponding quantifier in the input program, which could cause the translation to be unsound.

To avoid this unsoundness, our framework offers another parameter $\haveinv$ that is used to constrain the possible witnesses in the IVL-translation.
For any front-end type $t \in T_F$ and local variable $x \in \Variable$, $\haveinv(t,x)$ yields a syntactic boolean IVL expression that evaluates to \vcode{true} iff the value of variable $x$ in the IVL state has a corresponding front-end value of type $t$.
For example when $\trt{\fekeyword{Rat}}=\vkeyword{Real}$, we can define $\haveinv(\fekeyword{Rat},x) \triangleq ( \existstp{a}{\vkeyword{Int}} \existstp{b}{\vkeyword{Int}} b \ne 0 \wedge x=a/b )$. Using this framework parameter, we extend the translation of input assertions to all front-end assertions and obtain the assertion translation function $\tra{\cdot}$. The definition of $\tra{\existstp{x}{t}{A}}$ is as follows; that of the other cases is straightforward and is listed in \appendixref{appsec:framework}{A}.
{
    \small
    \[
    \tra{\existstp{x}{t} A} \triangleq \existstp{x}{\trt{t}} \haveinv(t,x) * \tra{A}
    \]
}

\begin{figure}
    \footnotesize
    \begin{align*}
        \trc{\fewhile{b}{I}{c}} &\triangleq \left( \vexhale{\tra{I}*\left(\trbe{b} \vee \neg\trbe{b}\right)}; \vhavoc{\wrc(c)}; \vinhale{\tra{I}*\neg\trbe{b}}, \phantom{\trcm{c}}\right.\\
        &\phantom{\triangleq} \left. \left\{ \vinhale{\tra{I}*\tra{b}}; \trcm{c}; \vexhale{\tra{I}*\left(\trbe{b} \vee \neg\trbe{b}\right)} \right\} \cup \trcext{c} \right)\\
        \trc{\fepar{P_1}{c_1}{Q_1}{P_2}{c_2}{Q_2}} &\triangleq \left( \vexhale{\tra{P_1*P_2}}; \vhavoc{\wrc(c_1) \cup \wrc(c_2)}; \vinhale{\tra{Q_1*Q_2}}, \phantom{\trcm{c}}\right.\\
        &\phantom{\triangleq} \left\{ \vinhale{\tra{P_1}}; \trcm{c_1}; \vexhale{\tra{Q_1}} \right\} \cup \\
        &\phantom{\triangleq} \left. \left\{ \vinhale{\tra{P_2}}; \trcm{c_2}; \vexhale{\tra{Q_2}} \right\} \cup \trcext{c_1} \cup \trcext{c_2} \right)
    \end{align*}
    \caption{The translation $\trc{\cdot}$ of front-end loops and parallel compositions. The full definition of $\trc{\cdot}$ is in \appendixref{appsec:framework}{A}. The condition $\trbe{b} \vee \neg\trbe{b}$ in the translation of loops excludes states in which the evaluation of $\trbe{b}$ fails (\eg due to a division by zero). Here we assume that the IVL supports (1) disjunction and negation of pure boolean expressions, and (2) directly using pure boolean expressions as assertions. It can therefore construct assertions including $\trbe{b} \vee \neg\trbe{b}$. $\vhavoc{V}$ is a shorthand for $\vhavoc{x_1}; \cdots; \vhavoc{x_n}$ for the finite set $V=\{x_1,\ldots,x_n\}$.}
    \label{fig:stmt-tr}
\end{figure}



Our framework translates loops and parallel compositions in an overapproximating way as verifiers like Viper do: the loop body or each thread in the parallel composition is translated together with its annotations as a separate program; we collect these programs to form a set of \emph{auxiliary IVL statements} $\trcext{c}$ for the front-end statement $c$. If $c$ does not contain loops or parallel compositions, $\trcext{c}$ is empty. In the main program, each statement $c$ is translated to $\trcm{c}$. If $c$ is a loop or parallel composition, $\trcm{c}$ is a sequence of exhale-havoc-inhale statements which first remove the resources specified by the annotations that are required to execute the block, then assign fresh values to all local variables that the execution of the block might modify, and finally add back the resources specified by the annotations that the execution of the block yields.
Given $\trcp{\cdot}$ that translates front-end primitive statements into (possibly composite) IVL statements, our framework thus defines the translation $\trc{c}$ of front-end composite statement $c$ as a pair of an IVL statement $\trcm{c}$ and a finite set of auxiliary IVL statements $\trcext{c}$. We give the definition of $\trc{\cdot}$ for loops and parallel compositions in \figref{fig:stmt-tr}. That for the other control structures is straightforward; the full definition of $\trc{\cdot}$ is in \appendixref{appsec:framework}{A}.

Finally, the provided parameters let us extend the encoding of front-end ghost heaps to an encoding $\trs{\cdot}$ of front-end ghost \emph{states}: $\trs{(s, \varphi)} \triangleq \left(\trst{s}, \trh{\varphi}\right)$, where the store encoding $\trst{s}$ is defined as:
{
    \small
    \begin{equation*}
    \trst{s}(x) = \left\{\begin{array}{ll}
        \text{undefined}, & s(x) \text{ is undefined}\\
        \trv{s(x)}, & \text{otherwise}
    \end{array}\right.
    \end{equation*}
}

\subsection{Requirements on the Framework Parameters}
\label{subsec:assumption}

In this subsection, we present the requirements on the framework parameters that guarantee the soundness of the front-end translation. Most of these requirements for the parameters from \figref{fig:fe-input} as well as the translations of types and values are standard and easy to satisfy; we list them in \appendixref{appsec:framework}{A}.

\reqref{req:st-tr} states the homomorphism property of the state encoding. We state it on the ghost heap encoding parameter $\trh{\cdot}$ of the framework rather than the derived ghost state encoding $\trs{\cdot}$; the homomorphism result of $\trs{\cdot}$ can be easily derived from that of $\trh{\cdot}$ in \reqref{req:st-tr}.
The second property here is a sanity requirement that prevents the heap encoding from introducing garbage that has no correspondence in the front-end; while in the diagram of \figref{fig:bsat}, backward satisfiability provides $\omega_A$, this property guarantees the existence of $\omega_F$ and thus helps complete the desired state split in the front-end.
Moreover, we require a third property, which allows us to support IDF\@.

\begin{requirement}\label{req:st-tr}
The encoding of ghost heaps $\trh{\cdot}$ must satisfy the following properties:
    \begin{itemize}
          \item $\trh{\cdot}$ preserves addition $\oplus$ when the sum of ghost heaps is defined in the front-end, \ie if $\varphi_1 \oplus \varphi_2$ is defined, then so is $\trh{\varphi_1} \oplus \trh{\varphi_2}$ and $\trh{\varphi_1} \oplus \trh{\varphi_2} = \trh{\varphi_1 \oplus \varphi_2}$.

        \item $\trh{\cdot}$ preserves subtraction in the following sense: if $\trh{\varphi} = \Phi_1 \oplus \trh{\varphi_2}$, then there exists some $\varphi_1$ such that $\Phi_1 \succeq \trh{\varphi_1}$ and $\varphi = \varphi_1 \oplus \varphi_2$, where $a \succeq b \triangleq (\existsntp{r}{a=b \oplus r})$.
        \item $\trh{\cdot}$ preserves stability. Specifically, $\stabilize{\trh{\varphi}}=\trh{\stabilize{\varphi}}$ for any front-end ghost heap $\varphi \in H_F$.
    \end{itemize}
\end{requirement}

With the parameters of the front-end language and the translation from \secref{subsec:syntax}, we formally define backward satisfiability for front-end assertions in \defref{def:backward-satisfiability}.

\begin{definition}[Backward satisfiability]
\label{def:backward-satisfiability}
A front-end assertion $A$ is \emph{backward satisfiable} \wrt{} the assertion translation $\tra{\cdot}$ iff: for any IVL state $\Omega$ that satisfies $\tra{A}$, and that is upper-bounded by the encoding of some front-end state $\omega_b$, \ie $\trs{\omega_b} \succeq \Omega$, there exists a front-end state $\omega$ which satisfies $A$ and \revision{$\Omega \oplus \core{\trs{\omega_b}} \succeq \trs{\omega}$}\footnote{\revision{The stronger statement $\Omega \succeq \trs{\omega}$ \hongyi{from \figref{fig:bsat}} might not be satisfied by some front-ends (\eg the one encoding lazy field creation in \secref{subsec:dfc-inst}) in the IDF algebra setting, as $\Omega$ might lose some pure information from $\omega_b$ which is necessary to construct $\omega$.}}.
\end{definition}

\reqref{req:be-asst} formalizes that the translations of boolean expressions, assertions, and statements must preserve the semantics in the front-end.
For assertions, we require in addition monotonicity of translated assertions, a property that is naturally satisfied in an affine logic\revision{, and that the translation does not add new free occurrences of $P$}. Together with backward satisfiability, it ensures that whenever the IVL satisfies a separating conjunction, the front-end can also satisfy it by redistributing the resources between the conjuncts.
%
We make the requirements only for input assertions in \reqref{req:be-asst}, and prove \lemref{thm:bsat-asst} as part of the framework that these properties hold for all front-end assertions.

\begin{requirement}\label{req:be-asst}
The translation functions for boolean expressions, input assertions, and primitive statements must satisfy the following properties:
    \begin{itemize}
\item
    The translation $\trbe{b}$ of any front-end boolean expression $b \in BE_F$ preserves the semantics of $b$, \ie for any front-end state $\omega \in \Sigma_F$, $\interpbe{b}(\omega)$ is defined iff $\trbe{b}(\trs{\omega})$ is defined, and they evaluate to the same truth value when both are defined.

\item
    The translation of input assertions $\trap{\cdot}$ must satisfy:
    \begin{itemize}
        \item $\trap{\cdot}$ preserves the semantics of \revision{all well-formed input assertions (\ie each $A$ such that $\wfa^{\Delta_F}(A)$)} in the front-end. Concretely, for any front-end state $\omega$ and any input assertion $A$, $\omega$ satisfies $A$ iff $\trs{\omega}$ satisfies $\trap{A}$.
        \item $\trap{A}$ is monotone for any input front-end assertion $A$. An assertion $Q$ is monotone iff for any states $\Omega$ and $\Omega'$, if $\Omega$ satisfies $Q$ and $\Omega' \succeq \Omega$, then $\Omega'$ also satisfies $Q$.
        \item \revision{$\trap{A}$ of any input front-end assertion $A$ does not contain any free occurrence of $P$.}
        \item All \revision{well-formed} input assertions are backward satisfiable \wrt{} $\trap{\cdot}$.
   \end{itemize}

\item
The translation of statements $\trcp{\cdot}$ preserves the semantics in the front-end, see formalization in \appendixref{appsec:framework}{A}.
    \end{itemize}
\end{requirement}

\begin{lemma}[Extending requirements to all front-end assertions]
\label{thm:bsat-asst}
If the translation $\tra{\cdot}$ is monotone\revision{, does not add new free occurrences of $P$,} and preserves the semantics for front-end assertions $A_1$ and $A_2$, and these two assertions are backward satisfiable \wrt{} $\tra{\cdot}$, then these properties are preserved by $\tra{\cdot}$ of
$A_1 * A_2$, 
$A_1 \vee A_2$, 
$A_1 \Rightarrow A_2$ (for a pure $A_1$), 
$\existstp{x}{t} A_1$\revision{, and
$\lfp{A_1}$
}.

As a corollary, $\tra{\cdot}$ is monotone and preserves the semantics for all \revision{well-formed} front-end assertions, and all \revision{well-formed} front-end assertions are backward satisfiable \wrt{} $\tra{\cdot}$.
\end{lemma}

In contrast, magic wands $\wand$, universal quantifications $\forall$, and iterated separating conjunctions $\oast$ do not necessarily preserve backward satisfiability of their components. For instance, assertions $W$, $P_1$, and $P_2$ from the two examples in \secref{subsec:unsoundness} are not backward satisfiable (as explained in the corresponding examples), even though their components are by \lemref{thm:bsat-asst}. For assertions containing these connectives, backward satisfiability needs to be proven without the help from \lemref{thm:bsat-asst}.


\subsection{Soundness}
\label{subsec:result}

The conditions presented in the previous subsection are sufficient to guarantee that the front-end translation described by a given instantiation of our framework is sound. In the following, we summarize the soundness argument. The full proof is in our \isabelle{} formalization.

One of our central lemmas shows that the state encoding preserves the validity of separating conjunctions if the first conjunct is monotone and the second is backward satisfiable. This property is essential to show that assertions retain the meaning, no matter whether they are assumed or asserted. It is, in fact, one of the main reasons we require assertions to be backward satisfiable:

\begin{lemma}[Translation of separating conjunctions]
\label{lem:sem-change-asst-tr}
    For any IVL semantic assertion $A$, define $\btr(A) \triangleq \{\omega \,|\, \trs{\omega} \in A\}$ as the backward translation of $A$, \ie the collection of all front-end states whose encodings satisfy $A$.

    For arbitrary IVL semantic assertions $A$ and $B$, $\btr{}(A) * \btr{}(B) \subseteq \btr{}(A*B)$.
    Furthermore, when $A$ is monotone, and $B$ is the translation of some backward satisfiable front-end assertion, the other direction also holds, \ie $\btr{}(A*B) \subseteq \btr{}(A)*\btr{}(B)$.
\end{lemma}

\lemref{lem:sem-change-asst-tr} allows us to prove \lemref{lem:exh-havoc-inh} which states that the exhale-havoc-inhale translation pattern that appears in the translation of both loops and parallel compositions (as in \figref{fig:stmt-tr}) is sound. The proof proceeds by applying the CSL frame rule on the CSL proof from the first property of \lemref{lem:exh-havoc-inh} with a carefully chosen frame assertion, followed by an application of the consequence rule where the implications at the pre- and postconditions in the consequence rule are justified by the two directions of implication in \lemref{lem:sem-change-asst-tr} respectively.

\begin{lemma}[Soundness of the exhale-havoc-inhale translation pattern]\label{lem:exh-havoc-inh}
    Assume that the framework parameters satisfy \reqref{req:st-tr} and \reqref{req:be-asst}, \revision{$P_0$ and $Q_0$ are well-formed,} and that the following properties hold:
    \begin{itemize}
        \item $\csl{\Delta_F}{\interpa{P_0}}{c}{\interpa{Q_0}}$. That is, the front-end loop or parallel composition block has a proof under CSL against the pre- and postconditions from the annotation of the block.
        \item $\axiomSem[\trctxt{\Delta_F}]{P}{\vexhale{\tra{P_0}}; \vhavoc{\wrc(c)}; \vinhale{\tra{Q_0}}}{Q}$. That is, the main program of the translation of the loop or parallel composition verifies against another given pair of pre- and postconditions in the IVL.
        \item The given IVL pre- and postconditions $P$ and $Q$ are monotone.
        \item All written variables of the loop or parallel composition block are declared, and thus included in the type context, \ie $\wrc(c) \subseteq \dom(\Delta_F)$.
    \end{itemize}
    Then $\csl{\Delta_F}{\btr{}(P)}{c}{\btr{}(Q)}$. That is, the front-end CSL proof for the block can be lifted to against the new pre- and postconditions backward translated from the IVL.
\end{lemma}

We can then prove our central soundness result stated as \thmref{thm:sound-tr-formal}. The proof proceeds by induction on the structure of front-end statements, where for the loop and parallel composition cases, we can apply \lemref{lem:exh-havoc-inh} with the IVL assertions $P$ and $Q$ in the lemma as the translation of the desired front-end pre- and postconditions $\tra{P}$ and $\tra{Q}$ and the first property derived by applying induction hypotheses on the axiomatic semantic proofs of the auxiliary IVL statements and CSL loop rule or parallel rule subsequently.

\begin{theorem}[Soundness]\label{thm:sound-tr-formal}
Assume that the framework parameters satisfy \reqref{req:st-tr} and \reqref{req:be-asst}.
    Let $c$ be a well-formed front-end composite statement \wrt{} type context $\Delta_F$, and $\trc{c}=(C_m,C_e)$ be the translation of $c$. Let $P$ and $Q$ be arbitrary \revision{well-formed} front-end composite assertions.
    Assume (1) $\axiomSem[\trctxt{\Delta_F}]{\tra{P}}{C_m}{\tra{Q}}$, and (2) \revision{any $C \in C_e$ verifies in the IVL, \ie} there exists an IVL semantic assertion $Q_C$ such that $\axiomSem[\trctxt{\Delta_F}]{\top}{C}{Q_C}$.
    Then $\csl{\Delta_F}{\interpa{P}}{c}{\interpa{Q}}$ holds in CSL.
\end{theorem}

This theorem guarantees that, whenever the IVL translation of a front-end program is correct then there exists a proof in CSL for the front-end program. To further connect this result to the operational semantics of the front-end language, we also prove the soundness of the CSL rules \wrt{} the small-step semantics. Our proof  adapts the proof from \citet{vafeiadisConcurrentSeparationLogic2011} to our IDF setting. The formal statement of this result is in \appendixref{appsec:framework}{A}.

\subsection{Generalization to State Encoding Relations}
\label{subsec:generalize-fw}

As we have seen in \secref{subsec:homomorphic-enc}, some front-ends might relate a single front-end state with multiple IVL states, as in the case of encoding duplicable permissions. To capture such encodings, we generalize the state encoding from a function $\trs{\cdot}$ to a binary relation $\rels{\cdot}{\cdot}$ between front-end states and IVL states and write $\rels{\omega}{\Omega}$ if IVL state $\Omega$ is \emph{one of the possible encodings} of front-end state $\omega$.

The generalized framework takes the same inputs as in \figref{fig:fe-input}, except that the heap encoding function $\trh{\cdot}$ is replaced by a binary relation $\relh{\cdot}{\cdot}$ that models the encoding of front-end ghost heaps into IVL heaps. The framework extends $\relh{\cdot}{\cdot}$ naturally to the binary relation $\rels{\cdot}{\cdot}$ between front-end and IVL \emph{states}: $\rels{(s, h)}{(S,H)} \triangleq (S= \trst{s} \wedge \relh{h}{H})$.

Most requirements of our framework are not affected by this generalization, but \reqref{req:st-tr} and \reqref{req:be-asst} need to be adapted:

\begin{grequirement}{req:st-tr}\label{greq:st-tr}
    The binary relation $\relh{\cdot}{\cdot}$ between front-end ghost heaps and IVL heaps need to satisfy the following properties.
    \begin{itemize}
        \item If $\varphi=\varphi_1 \oplus \varphi_2$ and $\relh{\varphi}{\Phi}$, then there exist $\Phi_1$ and $\Phi_2$ such that $\Phi=\Phi_1 \oplus \Phi_2$, $\relh{\varphi_1}{\Phi_1}$, and $\relh{\varphi_2}{\Phi_2}$.
        \item If $\relh{\varphi}{\Phi}$ and $\varphi_0=\varphi \oplus \varphi_1$, then there exists $\Phi_1$ such that $\relh{\varphi_1}{\Phi_1}$, $\Phi \oplus \Phi_1$ is defined, and $\relh{\varphi_0}{(\Phi \oplus \Phi_1)}$.
        \item For every front-end ghost heap $\varphi \in H_F$, there exists an IVL heap $\Phi \in \Sigma$ such that $\relh{\varphi}{\Phi}$.
        \item If $\relh{\varphi}{\Phi}$, then $\relh{\stabilize{\varphi}}{\stabilize{\Phi}}$.
        \item If $\relh{\stabilize{\varphi}}{\Phi'}$, then $\exists \Phi$ such that $\stabilize{\Phi}=\stabilize{\Phi'}$ and $\relh{\varphi}{\Phi}$.
    \end{itemize}
\end{grequirement}

Compared with \reqref{req:st-tr} in \secref{subsec:assumption}, the adaptation is in the following ways:
\begin{itemize}
\item The first property in \reqref{req:st-tr} (preservation of addition) is adjusted in a straightforward way.
\item The second property in \reqref{req:st-tr} (preservation of subtraction) is now subsumed by a generalized backward satisfiability, see below.
\item The second property here excludes certain ill-formed encodings by requiring that whenever a heap can be extended in the front-end, the corresponding extension is possible in the IVL (for instance, it is not possible to encode a fractional permission in the front-end to a full permission in the IVL). For the case where the heap encoding is a function, this property follows from the addition requirement and is, thus, not explicitly required in \reqref{req:st-tr}.
\item The third property here, intuitively, ensures that the encoding is non-degenerated.
\item The preservation of stability (\ie the last requirement in \reqref{req:st-tr}) is adapted to the last two requirements here.
\end{itemize}



We adjust the definition of backward satisfiability to \defref{def:gen-backward-satisfiability}. The definition coincides with the intuition in \figref{fig:bsat}, except that now $\omega_F$ may be encoded into an even larger $\Omega_F'$ rather than $\Omega_F$. This relaxation gives more flexibility with no affection on the soundness of the framework, intuitively because excess resources can always be discarded harmlessly in an affine SL.

\begin{definition}[Generalized backward satisfiability]
\label{def:gen-backward-satisfiability}
A front-end assertion $A$ has \emph{generalized backward satisfiability} \wrt{} the assertion translation $\tra{\cdot}$ iff: for any front-end state $\omega$ and IVL states $\Omega$, $\Omega_A$, and $\Omega_F$, if $\rels{\omega}{\Omega}$, $\Omega=\Omega_A \oplus \Omega_F$, and $\Omega_A$ satisfies $\tra{A}$, then there exist front-end states $\omega_A$ and $\omega_F$ and IVL state $\Omega_F'$ such that: $\omega=\omega_A \oplus \omega_F$, $\omega_A$ satisfies $A$, $\rels{\omega_F}{\Omega_F'}$, and $\Omega_F' \succeq \Omega_F$.
\end{definition}

With the generalized backward satisfiability notion, \reqref{req:be-asst} from \secref{subsec:assumption} needs only technical adaptations:

\begin{grequirement}{req:be-asst}\label{greq:be-asst}
The translation functions for boolean expressions, input assertions, and primitive statements must satisfy the following properties:
    \begin{itemize}
\item
    The translation $\trbe{b}$ of any front-end boolean expression $b \in BE_F$ preserves the semantics of $b$, \ie for any front-end state $\omega \in \Sigma_F$ and IVL state $\Omega \in \Sigma_V$ with $\rels{\omega}{\Omega}$, $\interpbe{b}(\omega)$ is defined iff $\trbe{b}(\Omega)$ is defined, and they evaluate to the same value when both are defined.

\item     The translation of input assertions $\trap{\cdot}$ must satisfy:
        \begin{itemize}
            \item $\trap{\cdot}$ preserves the semantics of all \revision{well-formed} input assertions in the front-end. Concretely, for any front-end state $\omega$ and IVL state $\Omega$ with $\rels{\omega}{\Omega}$, $\omega$ satisfies $A$ iff $\Omega$ satisfies $\trap{A}$.

            \item $\trap{A}$ is monotone for any $A \in A_{Fp}$.
            
            \item \revision{$\trap{A}$ of any input front-end assertion $A$ does not contain any free occurrence of $P$.}

            \item Any \revision{well-formed} $A \in A_{Fp}$ has generalized backward satisfiability \wrt{} $\trap{\cdot}$.
        \end{itemize}

\item
The translation of statements $\trcp{\cdot}$ preserves the semantics in the front-end.
        \end{itemize}
\end{grequirement}

To show that our framework for state encoding relations is indeed a generalization, we instantiate it, modeling the state encoding as a function, as stated in \thmref{thm:det-inst}.

\begin{theorem}[Generalization]\label{thm:det-inst}
Given an instantiation of the generalized framework where the state encoding is a function, if this instantiation satisfies all requirements from \secref{subsec:assumption} then it also satisfies the adapted requirements of the generalized framework.
\end{theorem}

Our main result is that any instantiation of the generalized framework that satisfies the adapted requirements is \emph{sound}. That is, \thmref{thm:sound-tr-formal}, with the adapted requirements, also holds in the generalized framework. 

The proof of the theorem requires several adaptations. 
In the generalized setting, we define the backward translation of IVL semantic assertion $P$ as $\btr{}(P) \triangleq \{\omega \,|\, \existsntp{\Omega \in P} \rels{\omega}{\Omega}\}$, and call $P$ well-formed iff $\btr{}(P)=\{\omega \,|\, \forall \Omega.\, \rels{\omega}{\Omega} \Rightarrow \Omega \in P\}$. The direction $\btr{}(A) * \btr{}(B) \subseteq \btr{}(A*B)$ in \lemref{lem:sem-change-asst-tr} does not hold for general $A$ and $B$ anymore, but requires $B$ to be well-formed. As a result, $Q_0$ being well-formed is needed for \lemref{lem:exh-havoc-inh}.
Despite these changes, we prove that all the assertions supported by a front-end instantiation of our generalized framework are well-formed as long as the instantiation satisfies all the requirements in \secref{subsec:generalize-fw}. As a result, the new premise in \lemref{lem:exh-havoc-inh} is always satisfied, and we get the same soundness result as \thmref{thm:sound-tr-formal} in the generalized framework.

The generalized framework provides an expressive foundation for  proving the soundness of practical translational verifiers. In particular, it reduces the soundness proof for state encodings to proving a small number of relatively simple requirements. We will demonstrate this expressiveness in \secref{subsec:immutable-heap-inst} by instantiating it with an encoding of duplicable permissions for immutable location.

\section{\revision{Applications of the Framework}}
\label{sec:instantiations}

In this section, we \revision{demonstrate the applicability of our framework by formalizing} three front-end translations with different state encodings into Viper, \revision{whose state model we formally define}
in \secref{subsec:viper-state}.
We first revisit the encoding from rational permission amount to reals (\secref{subsec:frac-perm-inst}),
\revision{and we show how to extend this encoding to support a backward-satisfiable fragment of magic wands used by Prusti~\cite{astrauskasLeveragingRustTypes2019}.}
We also translate a language with lazy field creation (\secref{subsec:dfc-inst}), and show an encoding of an immutable heap (\secref{subsec:immutable-heap-inst}).
\revision{We describe more encodings inspired by real-world front-ends (\secref{subsec:more-enc}). Finally, we discuss previously-reported soundness bugs in existing translational verifiers whose root causes manifest themselves as violations of our framework's requirements (\secref{subsec:real-world-unsoundness}).}

\revision{
In all the examples, our framework substantially simplifies the effort required for defining the translation and proving it sound: a formalization without our framework would essentially have to repeat the definitions and proofs of our main theorems,
in addition to proving the requirements of our framework.
Our framework also makes the soundness proof easy to maintain in practice, as it decomposes the \emph{global} soundness of a front-end translation into \emph{local} proof obligations: the translation of each input assertion must be backward satisfiable, and the translation of each primitive statement must be sound. This modularity keeps the proof robust under changes: adding a new input assertion (\eg the sound magic wand fragment in \secref{subsec:frac-perm-inst}) does not require revisiting the overall soundness argument, only proving the corresponding requirements for that assertion.
}

\subsection{Viper State Model}
\label{subsec:viper-state}

The formalization of Viper we use models the heap as a partial function from heap locations to values and the currently-owned permissions as a partial function from heap locations to real numbers, called \emph{permission mask}. A heap $h$ with a permission mask $\pi$ forms an IDF algebra with the following instantiations:
$\Sigma = \{ (h, \pi) \in (L \rightharpoonup V) \times (L \to [0,1]) \,|\, \forallntp{l} \pi(l) > 0 \Rightarrow h(l) \neq \bot \}$, where a heap location $l \in L$ is a pair of an address (modeled as natural number) and a field (identified by a unique name);
$\oplus$ adds up the permission amounts at each heap location and combines the values in the heap when the two heaps are compatible (the sum of permissions at each heap location does not exceed $1$, and the two heaps do not have different values at any heap location), \ie $(h_1,\pi_1) \oplus (h_2,\pi_2)$ is defined iff $\forallntp{l} \pi_1(l) + \pi_2(l) \le 1 \wedge (l \in \dom(h_1) \cap \dom(h_2) \Rightarrow h_1(l) = h_2(l))$, and equals to $(h_1 \cup h_2, \pi_1 + \pi_2)$ when it is defined;
$\core{\_}$ clears all the permissions, \ie $\core{(h,\pi)} \triangleq (h,\lambda l.\, 0)$;
$\stabilizeNa$ erases all values for heap locations to which no permission is held, \ie $\stabilize{(h,\pi)} \triangleq (\lambda l.\, \textsf{if } \pi(l)>0 \textsf{ then } h(l) \textsf{ else } \bot,\pi)$;
$\stableNa$ asserts whether every heap location storing a concrete value has a positive permission amount, \ie $\stable{(h,\pi)} \Leftrightarrow (\forallntp{l} h(l) \ne \bot \Rightarrow \pi(l) > 0)$;
$\unitNa$ erases the whole heap with the permissions, \ie $\unit{(h,\pi)}=(\lambda l.\, \bot, \lambda l.\, 0)$;

A Viper state is thus an IVL state where the customized heap is instantiated with a heap with a permission mask, \ie $\Sigma_V=(\Variable \rightharpoonup V) \times \Sigma$ with $\Sigma$ defined above.

\subsection{Encoding Rational Permission Amounts into Real Permission Amounts}
\label{subsec:frac-perm-inst}

In \secref{subsec:unsoundness}, we showed that the common encoding of fractional permissions with rational permission amounts into real permission amounts is generally unsound. Our example there made use of magic wands. In this subsection, we show that such an encoding is sound for a front-end with a more restricted assertion language. 

To focus on the essentials, we assume that the front-end language is very similar to our IVL, Viper. Both have the same types (\eg for booleans, integers, and references), values (\eg $\fekeyword{null}$), and expressions (except that the front-end uses rational arithmetic whereas the IVL uses reals). Among the primitive 
statements, local variable assignment and field assignment are common to front-end language and IVL; moreover, the front-end supports allocation and de-allocation, whose translation is discussed below.

We provide the details of the instantiation of our framework in \appendixref{appsubsec:frac}{B.1} and focus on the most interesting aspects here.
The front-end models the program heap as a partial map from heap locations (as pairs of natural number identifiers and string-named fields, which is the same as in Viper) to values, and uses rational permission amounts for its ghost heap, \ie $H_P=\{h_p:L \rightharpoonup_{\textsf{fin}} V_F\}$
where $\rightharpoonup_{\textsf{fin}}$ represents a finite partial mapping
, and $H_F=\{(h,\pi) \in (L \rightharpoonup V_F) \times (L \rightarrow [0,1] \cap \mathbb{Q}) \,|\, \forallntp{l} \pi(l)>0 \Rightarrow h(l) \ne \bot\}$. The embedding of $H_P$ onto $H_F$ keeps the partial heap, and assigns full permissions to heap locations that contain values and zero permissions otherwise: $\embedh{}(h_p)=(h_p, \lambda l.\, \textsf{ if } h_p(l)\ne\bot \textsf{ then } 1 \textsf{ else } 0)$.

Due to the similarity of the front-end language and IVL, the translation functions for types, values, ghost heaps, and expressions are straightforward, except that the type $\fekeyword{Rat}$ is translated to $\vkeyword{Real}$. $\haveinv{}(t,x)$ is thus non-trivial only for $t=\fekeyword{Rat}$ as $\haveinv{}(\fekeyword{Rat},x) \triangleq \existstp{a}{\vkeyword{Int}} \existstp{b}{\vkeyword{Int}} b \ne 0 \wedge x=a/b$.

The input assertion $\feaccpred{x}{f}{p}$ is translated unchanged to $\vaccpred{x}{f}{p}$. Preservation of semantics of the translation and monotonicity of the translated assertion are straightforward.
Backward satisfiability is also intuitive: if $\tre{p}$\footnote{We use $\tre{\cdot}$ to denote the translation of general expressions (including non-boolean ones). It is not a parameter of our framework, but framework parameters can use it (like any other mathematical function).} evaluates to $v$ under some $\Omega$ with $\trs{\omega_b} \succeq \Omega$, then it also evaluates to $v$ under $\trs{\omega_b}$ (adding resources cannot change a defined evaluation), and hence to $v$ under $\omega_b$ in the front-end, so $v$ is rational. The front-end state holding just $v$ permission to $x.f$ (and only the information needed to evaluate $x$ and $p$) is therefore the smallest state satisfying $\feaccpred{x}{f}{p}$, and its encoding is smaller than or equal to $\Omega$.

The translations of variable assignment and field assignment are straightforward. Allocation and de-allocation are translated as follows:
{
    \vspace{-0.5mm}
    \small
    \begin{align*}
        \trcp{\fealloc{x}{f_1,\ldots,f_n}} &\triangleq \vhavoc{x}; \vinhale{\vaccpred{x}{f_1}{1} * \cdots * \vaccpred{x}{f_n}{1}} \\
        \trcp{\fefree{x.f}} &\triangleq \vexhale{\vaccpred{x}{f}{1}}
    \end{align*}
}
Intuitively, the translation of allocation picks non-deterministically a fresh heap location identifier ($\vhavoc{x}$) and provides the necessary permissions for the allocated heap locations; the translation of de-allocation removes the permission to the deleted heap location. 

We prove in \isabelle{} that this instantiation satisfies all requirements of our framework.
This instantiation shows that translating rational permission amounts as reals is not per se unsound. If we had added \revision{general} magic wands as input assertions to our instantiation (as in \figref{fig:rational-real-wand-unsound} in \secref{subsec:unsoundness}) then the proof would have failed because \reqref{req:be-asst} would not hold.
\revision{Nevertheless, our framework can still support a restricted fragment of magic wands.}

\paragraph{\revision{A sound fragment of magic wands.}}
\revision{
In practice, the magic wands used by existing verifiers are often far more limited.
Prusti~\cite{astrauskasLeveragingRustTypes2019}, for instance, uses only wands whose two sides are \emph{binary assertions} (\ie{} never use fractional permissions), and that do not access rational-valued fields.
Such assertions are given by the following grammar
(we further check that the fields they read do not have the type $\fekeyword{Rat}$ with the framework's well-formedness predicate $\wfa^{\Delta_F}$),
where $e \in BE_F$ and $t \in T_F$:
{
    \small
  \[
  A_b \Coloneqq \;
  \feaccpred{x}{f}{1} \mid
  e \mid
  A_b * A_b \mid
  A_b \vee A_b \mid
  e \Rightarrow A_b \mid
  \existstp{x}{t} A_b
  \]
}
}
\revision{
\noindent
Each such assertion is translated like the corresponding case of $\tra{\cdot}$ in \appendixref{appsec:framework}{A}; we write $\trab{\cdot}$ for this translation. We then add to the front-end's input assertions all wands $L \wand R$ whose two sides are binary assertions that do not access rational fields, with the standard interpretation and $\trap{L \wand R} \triangleq \trab{L} \wand \trab{R}$.
We prove in \isabelle{} that $\trap{L \wand R}$ satisfies \reqref{req:be-asst}, so this fragment of magic wands is captured by our framework and, thus, sound.
}

\subsection{Encoding Lazy Field Creation}
\label{subsec:dfc-inst}

In \secref{sec:intro}, we showed an encoding of a Python-like front-end language with lazy field creation into an IVL with static field declarations. In this subsection, we define a front-end that formalizes the lazy field creation feature, and a translation into Viper. Our framework is able to capture this translation to prove its soundness. This instantiation shows, in particular, that our framework is able to express translations with different heap structures (each front-end field is encoded as two IVL fields) and custom proof rules (as we explain below).

We assume a front-end similar to the one from \secref{subsec:frac-perm-inst}. Its logic reasons about lazy field creation as follows: 
It governs access to a location $x.f$ using the standard accessibility predicate $\feaccpred{x}{f}{p}$. We assume that the front-end can determine statically which fields are \emph{potentially} added to an object, for instance, by using a type system such as Python's MyPy. Allocation provides permission to all those fields and sets their value to $\fenone{}$ to indicate that this field has not been created yet. Field updates require permission to access the field, as usual. The first assignment to a field creates the field, which is reflected by replacing $\fenone{}$ with the  assigned value. Finally, field read requires permission and asserts that the field value is different from $\fenone{}$, as attempting to read a field before it has been added triggers a runtime error. Since $\fenone{}$ is not part of the surface syntax, the frontend provides pure input assertion $\initpred{x}{f}$ and $\uninitpred{x}{f}$ to denote $x.f\neq\fenone{}$ and $x.f=\fenone{}$, respectively.

To reflect this logic, the front-end provides the designated value $\fenone{}$ and uses it in its heap model:
$H_P=\{h_p:L \rightharpoonup_{\textsf{fin}} V_F \cup \{\fenone{}\}\}$ and $H_F=\{(h,\pi) \in (L \rightharpoonup V_F \cup \{\fenone{}\}) \times (L \rightarrow [0,1] \cap \mathbb{Q}) \,|\, \forallntp{l} \pi(l)>0 \Rightarrow h(l) \ne \bot\}$. The evaluation of a heap access expression $x.f$ is defined only if its value is different from $\fenone{}$. Other than that, the evaluation of expressions is the same as in \secref{subsec:frac-perm-inst}.

Our front-end supports the input assertions $\feaccpred{x}{f}{p}$, $\initpred{x}{f}$, and $\uninitpred{x}{f}$. Compared with the interpretation of $\feaccpred{x}{f}{p}$ in the previous subsection, the interpretation here additionally accepts the case where the field is missing, but excludes the case where $p$ evaluates to $0$. The interpretation of $\initpred{x}{f}$ and $\uninitpred{x}{f}$ is straightforward. The concrete definition of $\interpana$ for the three input assertions is shown in \appendixref{appsubsec:dfc}{B.2}. 

Local variable assignments, allocations, and de-allocations have the same form as in the previous subsection. For simplicity, we restrict field assignments to the form $\fefieldassign{x}{f}{v}$, where $v \in \Variable$ is a local variable. 
The CSL rules for the primitive statements are the same as before, and so are the operational semantic reduction rules for local variable and field assignments and de-allocations. For an allocation $\fealloc{x}{f_1,\ldots,f_n}$, the allocated locations are initialized to $\fenone{}$ rather than arbitrary concrete values. The formal reduction rule is shown in \appendixref{appsubsec:dfc}{B.2}.

The most interesting aspect of this front-end is its state encoding.
Each front-end field $f$ is encoded by \emph{two} IVL fields: The boolean field $\addedf{}$ indicates whether the field $f$ has been added, whereas the field $\ivlf{}$ stores the translated value of the front-end heap location. That is, our encoding reflects the cases where a field has been added or not as follows ($v$ is different from $\fenone{}$). 

\begin{center}
{\small
    \begin{tabular}{cc}
        $a.f \stackrel{p}{\mapsto} \fenone$ for $p \in [0,1]$ & $a.\addedf{} \stackrel{p}{\mapsto} \vfalse$, $a.\ivlf{} \stackrel{0}{\mapsto} \bot$\\[1mm]
        $a.f \stackrel{p}{\mapsto} v$ for $p \in [0,1]$ & 
$a.\addedf{} \stackrel{p}{\mapsto} \vtrue$, $a.\ivlf{} \stackrel{p}{\mapsto} \trv{v}$
    \end{tabular}
}
\end{center}

Expressions are translated identically, except that every field name $f$ in field access expressions is replaced by $\ivlf{}$. 
The translation of input assertions needs to reflect the above state encoding. In particular, $\tra{\feaccpred{x}{f}{p}}$ denotes permission to $\addedf{}$ and, in case the field has been added, also permission to $\ivlf{}$:
{
    \vspace{-0.5mm}
    \small
\[
\trap{\feaccpred{x}{f}{p}}\triangleq
\tre{p} > 0 *
\vaccpred{\tre{x}}{\addedf{}}{\tre{p}} * (
\tre{x}.\addedf{} = \vtrue \Rightarrow \vaccpred{\tre{x}}{\ivlf{}}{\tre{p}}
)
\]}

Field assignments are translated as follows:
{
    \small
    \begin{align*}
        \trcp{\fefieldassign{x}{f}{v}} &\triangleq \vif{\tre{x}.\addedf{}=\vfalse}{\vinhale{\vaccpred{\tre{x}}{\ivlf{}}{1}}}{\vskipp};\\
        &\phantom{\triangleq} \vfieldassign{\tre{x}}{\addedf{}}{\vtrue};
        \vfieldassign{\tre{x}}{\ivlf{}}{v}
    \end{align*}
}
Intuitively, it uses the field $\tre{x}.\addedf{}$ to check whether this assignment dynamically creates the field. If so, it adds full permission to the field $\tre{x}.\ivlf{}$, which enables future read and write accesses. Moreover, it records in $\tre{x}.\addedf{}$ that the field has been created (in either case, as this is a monotonic update), and assigns the value to 
$\tre{x}.\ivlf{}$. The translation of the other primitive statements is straightforward; the full definition is presented in \appendixref{appsubsec:dfc}{B.2}. 

The front-end translation described above satisfies all requirements of our framework (for state encoding functions) and is, thus, guaranteed to be sound.

\subsection{Encoding Immutable Heaps}
\label{subsec:immutable-heap-inst}

Our final instantiation illustrates how our framework supports duplicable resources into an IVL that offers only (non-duplicable) fractional resources. Duplicable resources have many applications, for instance, as permission to acquire a lock or to represent monotonic information. In this subsection, we use access to immutable data as an example for duplicable resources.

To avoid the complications of initializing immutable fields, we assume a front-end here, where immutable data is stored in a separate heap. Field accesses select the heap using different dereferencing operators, $x.f$ for the mutable heap and $\imacc{x}{f}$ for the immutable heap.\footnote{Note that both heaps share the same identifiers, so a field $f$ may occur in both heaps, and a syntactic distinction of the two accesses is necessary.} 
Allocation in the mutable heap works as usual, whereas allocation in the immutable heap takes as arguments the initial values of the fields and initializes them. Hence, regular allocation provides the usual ownership of the allocated locations, donated again by 
$\feaccpred{x}{f}{p}$, whereas immutable allocation provides a \emph{duplicable} resource $\iminitpred{x}{f}$, which permits read access, but not assignments. 

To reflect these ideas, our front-end is similar to the one in \secref{subsec:frac-perm-inst}, but has two heaps: 
$H_P=\{(h,h_{im}) \in (L \rightharpoonup V_F) \times (L \rightharpoonup V_F)\}$, where the first component is the mutable partial heap and the second one is the immutable heap. The ghost heap set is $H_F=\{(h,\pi,h_{im}) \in (L \rightharpoonup V_F) \times (L \to [0,1] \cap \mathbb Q) \times (L \rightharpoonup V_F) \,|\, \forallntp{l} \pi(l)>0 \Rightarrow h(l) \ne \bot\}$. The embedding function $\embedh{}$ maps the immutable heap $h_{im}$ identically and the mutable heap $h$ as to a heap $(h,\pi)$ with rational permissions. The resource of each heap location in the immutable heap is stable and duplicable, \ie the immutable heap in $H_F$ is instantiated as an IDF algebra with the following definitions of the operations:
\begin{itemize}
    \item $h_1 \oplus h_2$ is defined iff $\forall l \in L$, if both $h_1(l)$ and $h_2(l)$ are defined then $h_1(l)=h_2(l)$. In this case, $(h_1 \oplus h_2)(l) \triangleq \textsf{ if } h_1(l) \text{ is defined then } h_1(l) \text{ else } h_2(l)$.
    \item $\core{h} = \stabilize{h} \triangleq h$, and $\stable{h}$ holds for any $h$.
    \item $\unit{h} \triangleq \emp{}$, where $\emp{}$ is the empty heap $\lambda l.\, \bot$.
\end{itemize}

The front-end supports the same primitive statements as in \secref{subsec:frac-perm-inst}. In addition, the statement $\feinit{x}{f}{v}$ allocates and initializes an immutable location  $\imacc{x}{f}$.
The operational semantic reduction rule and CSL rule for this statement are straightforward; we list them in \appendixref{appsubsec:immutable-heap}{B.3}.

Since our IVL does not provide duplicable resources, we encode them as arbitrary positive fractional permission amounts. As a result, the representation for a duplicable front-end resource is not unique. Our generalized framework allows us to express such state encoding relations.

The encoding of front-end ghost heaps is modeled as a binary relation.
Both front-end heaps are encoded into a single IVL heap. To disambiguate accesses, we use a map $f_h$ that maps each front-end field $f$ in the mutable heap to the IVL field $hf$; the map $f_v$ analogously maps each front-end field $g$ in the immutable heap to the IVL field $vg$. The ranges of $f_h$ and $f_v$ are disjoint. The encoding of the mutable heap is the same as in 
\secref{subsec:frac-perm-inst} (except that here field names are mapped). 
The (allocated) heap locations in the immutable heap are encoded to having \emph{any} permission $p \in (0,1)$; that is, the state relation permits \emph{all} such amounts $p$. Un-allocated locations have zero permission and no value.

To formalize the state encoding relation, we need to divide the IVL heap into the parts corresponding to the mutable and immutable front-end heaps. In those parts, we then undo the field name mappings $f_h$ and $f_v$ so that we can relate them to the front-end fields.
Concretely, we define $H|_{f_h}$ to be the IVL sub-heap constructed by taking only locations with field names in the range of $f_h$; we construct $H|_{f_v}$ analogously. $H_{\textsf{rem}}$ denotes the remaining part of $H$ after removing $H|_{f_h}$ and $H|_{f_v}$. $\Pi|_{f_h}$, $\Pi|_{f_v}$, and $\Pi_{\textsf{rem}}$ are defined similarly for the permission mask $\Pi$ (while ``removing'' means setting the permission to each location in the corresponding part to zero). Now we can define the state relation, by reusing the ghost heap encoding $\trh{\cdot}$ from \secref{subsec:frac-perm-inst} for the mutable heap:
{
    \small
    \begin{align*}
        \relh{(h,\pi,h_{im})}{(H,\Pi)} &\triangleq \left(H|_{f_h}, \Pi|_{f_h}\right)=\trh{(h,\pi)} \wedge
        H_{\textsf{rem}} = \lambda l.\, \bot \wedge \Pi_{\textsf{rem}} = \lambda l.\, 0 \wedge
        (\forall l. \\
        &\phantom{\triangleq}
        (h_{im}(l) = \bot \Rightarrow (H|_{f_v}(l)=\bot \wedge \Pi|_{f_v}(l)=0)) \wedge{} \\
        &\phantom{\triangleq} (h_{im}(l) = v \Rightarrow (H|_{f_v}(l)= \trv{v} \wedge \Pi|_{f_v}(l) \in (0,1))))
    \end{align*}
}

Heap access expressions are translated accordingly: the field names for all accesses to the mutable heap (of the form $x.f$) are translated by $f_h$, while those for the accesses to the immutable heap (of the form $\imacc{x}{f}$) are translated by $f_v$. The translation for the input assertions are as follows, where $\vaccpred{x}{f}{\_}$ represents ownership of an arbitrary positive fractional permission amount for $x.f$.
{
    \small
    \begin{align*}
        \trap{\feaccpred{x}{f}{p}} &\triangleq \vaccpred{\tre{x}}{f_h(f)}{\tre{p}} &
        \trap{\iminitpred{x}{f}} &\triangleq \vaccpred{\tre{x}}{f_v(f)}{\_}
    \end{align*}
}

The translation of local variable assignments, allocations in the mutable heap, and de-allocations is similar to that in \secref{subsec:frac-perm-inst}; we show them in \appendixref{appsubsec:immutable-heap}{B.3}. The translation of allocation in the immutable heap assigns a fresh heap location, adds the resource corresponding to $\iminitpred{x}{f}$ to the state, and assumes that the new location has its initial value:
{
    \small
\[
\trcp{\feinit{x}{f}{v}} \triangleq \vhavoc{x}; \vinhale{\vaccpred{x}{f_v(f)}{\_} * x.f_v(f)=v}
\]
}

We proved in \isabelle{} that the instantiation of our \emph{generalized} framework for immutable data satisfies all requirements and is, thus, guaranteed to be sound.
This instantiation also demonstrates that our generalized framework is sufficiently expressive to handle complex, but common state encodings such as translating duplicable resources to non-duplicable fractional resources.

\subsection{\revision{Additional Encodings Inspired by Real-World Front-Ends}}
\label{subsec:more-enc}

\revision{
Beyond the three instantiations formalized above, we believe that our framework can capture many other front-ends. To illustrate this point, we outline three further state encodings, inspired by real-world front-ends, that also fit our framework:}

\revision{
\textit{The Pancake verifier's state model~\cite{zhao2025verifyingdevicedriverspancake}:} The front-end heap is a fixed-size array of cells, encoded into a Viper sequence \vcode{heap} of references, where the cell at offset \fecode{i} corresponds to the heap location \vcode{heap[i]} refers to. It uses the same fractional permission model as Viper.
}

\revision{
\textit{The block-offset memory model for C, adapted from \citet{leroy2009formal}:} C memory is modeled as an unbounded but finite set of blocks, each with a unique identifier, a linear array of cells (with binary permissions) indexed by their offset, and a fixed size.
The encoding maps each cell (identified by its block and offset) to a unique heap location storing its value in a \vcode{val} field, and each block to a heap location whose \vcode{size} field stores the block's size.
}

\revision{
\textit{The object-oriented memory model from \citet{loow2025compositional}:} The front-end heap consists of an unbounded but finite set of objects, each with a unique identifier, a list of cells (with binary permissions) named by field, and a domain (the set of its field names).
The encoding maps each object to a unique heap location, each cell to the field of that location with the same name, and stores the object's domain in an additional field.
}

\revision{In all three instantiations, the homomorphism requirement holds:
Two heaps in the Pancake state model must agree on the array size and the content of each cell, allowing the combination of the encodings; the permissions are simply added for each cell.
In the other two models, two states with the same identifier must agree on their shared contents and have disjoint domains on the non-duplicable contents, so their encodings combine: two C blocks must agree on their size and two objects on their domain; blocks or objects from two states must own disjoint cells.
Backward satisfiability of the input assertions (\eg fractional ownership of a cell, or a block's size) is likewise straightforward.
}

\subsection{\revision{Identifying Unsoundness in Real-World Front-End Translations}}
\label{subsec:real-world-unsoundness}

\revision{
In addition to proving existing front-ends sound, our framework captures common sources of unsoundness in existing verifiers: the root causes of several previously-reported soundness bugs correspond to violations of our requirements, showing that checking these requirements could have caught them. For example, issues \#1189~\cite{prustiissue1189} and \#1518~\cite{prustiissue1518} in Prusti~\cite{astrauskasLeveragingRustTypes2019} are caused by passing the same shared reference as two arguments to a method call, which creates two independent fractional permissions in the IVL, whereas the front-end state model forbids such a split, which thus violates the homomorphism requirement in \reqref{req:st-tr}. Violations of \reqref{req:be-asst} underlie issue \#955~\cite{vercorsissue955} (an unsound quantifier-simplification rewrite) in VerCors~\cite{blomVerCorsToolSet2017}, issue \#1501~\cite{prustiissue1501} (wrong encoding of \fecode{matches!} inside assertions) in Prusti, and issue \#281~\cite{naginiissue281} (unsound bool/object round-trip cast) in Nagini~\cite{eilersNaginiStaticVerifier2018}.
Violations of requirements in \appendixref{appsec:framework}{A} underlie further bugs: soundness of the front-end SL rules is violated by issue \#703~\cite{vercorsissue703} (an unreachable send is treated as transferring permission) in VerCors, injectivity of value translation by issue \#1357~\cite{vercorsissue1357} (\fecode{int} and \fecode{short} are both mapped to the same IVL type) in VerCors, and soundness of the statement translation by issue \#161~\cite{naginiissue161} (static field updates are not propagated) in Nagini.
}


\section{Related Work}
\label{sec:related-work}

Our work was inspired by \citet{dardinierFormalFoundationsTranslational2025}'s framework for proving the soundness of SL-based translational verifiers. However, their framework requires the front-end and the IVL to have the same state model, which is not the case in existing translational SL-verifiers, such as
Gobra,
Nagini~\cite{eilersNaginiStaticVerifier2018},
a Pancake verifier~\cite{zhao2025verifyingdevicedriverspancake},
Prusti, and
VeriFast. In contrast, our framework supports complex state encodings, both in terms of the representation of values and of SL resources.

Several existing works target the soundness of front-end translations. For instance, \citet{SummersM20} and \citet{WolfSM22} sketch soundness proofs for the Viper translations of the RSL weak memory logic~\cite{VafeiadisN13} and the TaDA logic~\cite{PintoDG14}, respectively. Both involve non-standard SL resources, for which our framework provides a formal foundation.

Instead of translating state models to the IVL, Gillian uses an IVL and verification back-end that are parametric in the state model. This simplifies front-end soundness proofs but requires the back-end to be sound for all instantiations. \citet{loow2025compositional} present a framework for such proofs in Rocq, which is parametric in the concrete and symbolic state models, with sufficient conditions for a sound instantiation. However, unlike our work, it includes no statement or assertion translations and supports neither loops nor concurrency, so it is unclear whether it can capture realistic front-end translations. Moreover, it does not facilitate soundness proofs for IVLs that are not parametric in the state model, such as Raven, VeriFast, and Viper. 

Front-end translations to IVLs not based on SL~\cite{VogelsJP09,Backes2011,Herms13,Fortin13a,THSem} involves generally much simpler state encodings. In fact, the existing unsound encodings from \secref{subsec:homomorphism} and \secref{subsec:unsoundness} as well as our sufficient soundness conditions are all concerned with the representation of SL resources. 

Some existing approaches target the soundness of SL-based verification back-ends.
\citet{THSem} present a technique for automatically producing certificates for successful verification runs in Viper. 
\citet{zimmermanSoundGradualVerification2024} present a formalization of a variant of Viper's symbolic execution back-end for gradual verification. \citet{jacobsFeatherweightVeriFast2015} formalize and prove sound a simplified version of the symbolic execution back-end of VeriFast~\cite{JacobsSPVPP11} in Rocq. 
These back-end soundness proofs can be combined with our framework's front-end proofs for an overall soundness guarantee.

\section{Conclusion and Future Work}
\label{sec:conclusion}

We presented the first framework, formalized in \isabelle{}, for proving the soundness of the front-ends of translational separation logic verifiers with complex state encodings, which are pervasive in existing verifiers. Our sufficient soundness conditions greatly simplify soundness proofs and help uncover flawed encodings quickly, as we illustrate on several common encodings. 

As future work, we plan to extend the framework to support additional concurrency primitives (\eg threads and locks) and to apply the framework to prove soundness of existing front-ends.

\section*{Data Availability Statement}
All technical results presented in this paper have been formalized and proved in \isabelle{}. The formalization is publicly available in our artifact~\cite{ling_artifact}. The latest version is available at \url{https://doi.org/10.5281/zenodo.21509501}.


{
\interlinepenalty=10000
\bibliography{references}
}

\appendix

\ifextended
\clearpage
\section{Full Definitions and Semantics in the Framework}
\label{appsec:framework}

The axioms for IDF algebra is listed in \figref{fig:idf-axiom}.

\begin{figure}[ht]
    \small
    \begin{mathpar}
        a \oplus b = b \oplus a \and
        a \oplus (b \oplus c) = (a \oplus b) \oplus c \and
        c = a \oplus b \wedge c = c \oplus c \Rightarrow a = a \oplus a \and
        x = x \oplus \core{x} \and
        \core{x} = \core{x} \oplus \core{x} \and
        x = x \oplus c \Rightarrow \core{x} \succeq c \and
        \core{a \oplus b} = \core{a} \oplus \core{b} \and
        a = b \oplus x \wedge a = b \oplus y \wedge \core{x} = \core{y} \Rightarrow x = y \and
        \stable{\omega} \Rightarrow \omega = \stabilize{\omega} \and
        \stable{\stabilize{\omega}} \and
        \stabilize{a \oplus b} = \stabilize{a} \oplus \stabilize{b} \and
        x = \stabilize{x} \oplus \core{x} \and
        a = b \oplus \unit{c} \Rightarrow a = b \and
        a = a \oplus \unit{a} \and
        \core{\unit{a}} = \unit{a} \and
        \stable{\unit{a}}
    \end{mathpar}
    \caption{Axioms for the IDF algebra $(\Sigma,\oplus, \core{\_}, \stableNa, \stabilizeNa, \unitNa)$. Define $\omega' \succeq \omega \triangleq (\exists r. \omega'=\omega \oplus r)$.}
    \label{fig:idf-axiom}
\end{figure}

The axiomatic semantic rules for IVL statements $\vhavoc{x}$, $\vinhale{A}$, $\vexhale{A}$, $C_1;C_2$, and $\vif{b}{C_1}{C_2}$ are listed in \figref{fig:ax-sem-vpr}.

\begin{figure}[ht]
    \footnotesize
    \begin{mathpar}
        \inferhref{InhaleAx}{rule:InhaleAx}
        {\selfFraming{P} \\
        \framedBy{A}{P}}
        {\axiomSem{P}{\vinhale{A}}{P * A}}

        \and

        \inferhref{ExhaleAx}{rule:ExhaleAx}
        {\selfFraming{P} \\
        P \subseteq Q * A \\
        \selfFraming{Q}}
        {\axiomSem{P}{\vexhale{A}}{Q}}

        \and

        \inferhref{HavocAx}{rule:HavocAx}
        {\selfFraming{P} \\
        \Delta(x) = \tau}
        {\axiomSem{P}{\vhavoc{x}}{\exists x \in \tau \ldotp P}}

        \and

        \inferhref{SeqAx}{rule:SeqAx}
        {\axiomSem{P}{C_1}{R} \\
        \axiomSem{R}{C_2}{Q}}
        {\axiomSem{P}{C_1;C_2}{Q}}

        \and

        \inferhref{IfAx}{rule:IfAx}
        {\selfFraming{P} \\
        \framedBy{P}{b} \\
        \axiomSem{P * b}{C_1}{B_1} \\
        \axiomSem{P * \lnot b}{C_2}{B_2}}
        {\axiomSem{P}{\vif{b}{C_1}{C_2}}{B_1 \lor B_2}}
    \end{mathpar}
    \caption{Axiomatic semantics for the IVL statements used in our framework.}
    \label{fig:ax-sem-vpr}
\end{figure}

\revision{The extension of $\wfa^{\Delta_F}$ for front-end assertions is defined via an auxiliary function $\wfaaux{A}{B}$, where $B$ is a boolean indicator of whether $A$ is in the scope of any $\mu$ binder. $\wfaaux{A}{B}$ is defined as follows, where $\Delta_F[x:=t]$ represents the type context derived from $\Delta_F$ by replacing the type of local variable $x$ to $t$, and $\wfa^{\Delta_F}(A) \triangleq \wfaaux{A}{\text{false}}$.
{
    \small
    \begin{align*}
        \wfaaux{A_p}{B} &\triangleq \wfa^{\Delta_F}(A_p)\\
        \wfaaux{A_1 * A_2}{B} &\triangleq \wfaaux{A_1}{B} \wedge \wfaaux{A_2}{B}\\
        \wfaaux{A_1 \vee A_2}{B} &\triangleq \wfaaux{A_1}{B} \wedge \wfa{A_2}{B}\\
        \wfaaux{\existstp{x}{t} A}{B} &\triangleq \wfaauxdelta{\Delta_F[x := t]}{A}{B}\\
        \wfaaux{b \Rightarrow A}{B} &\triangleq \wfaaux{A}{B}\\
        \wfaaux{b}{B} &\triangleq \text{true}\\
        \wfaaux{P}{B} &\triangleq B\\
        \wfaaux{\lfp{A}}{B} &\triangleq \wfaaux{A}{\text{true}}
    \end{align*}
}
}

\revision{To define the extension of interpretation function $\interpa{\cdot}$ and free variable function $\fva$ for front-end assertions, we first define auxiliary interpretation $\interpaaux{\cdot}{\Gamma}$ parameterized by the interpretation $\Gamma$ for predicate symbol $P$ and auxiliary function $\fvaaux{\cdot}{F}$ parameterized by the free variable set $F$ of predicate symbol $P$ as follows, where $\omega[v/x]$ represents the resulting state of overwriting the value of local variable $x$ in $\omega$ to $v$, and $\LFP{S}{f(S)} \triangleq \bigcap_{S : f(S) \subseteq S}S$ is the least fixed point of $f$ when it monotonely increases on $S$, which we prove is the case for $\interpaaux{\cdot}{\Gamma}$ and $\fvaaux{A}{F}$. Then $\interpa{A} \triangleq \interpaaux{A}{\varnothing}$ and $\fva(A) \triangleq \fvaaux{A}{\varnothing}$.}

\revision{
{
    \small
    \begin{align*}
        \interpaaux{A_1 * A_2}{\Gamma} &\triangleq \{\omega \,|\, \existsntp{\omega_1 \in \interpaaux{A_1}{\Gamma},\omega_2 \in \interpaaux{A_2}{\Gamma}} \omega = \omega_1 \oplus \omega_2\}\\
        \interpaaux{A_1 \vee A_2}{\Gamma} &\triangleq \interpaaux{A_1}{\Gamma} \cup \interpaaux{A_2}{\Gamma}\\
        \interpaaux{\existstp{x}{t} A}{\Gamma} &\triangleq \{\omega \,|\, \existsntp{v} \tp{}(v)=t \wedge \omega[v/x] \in \interpaauxdelta{\Delta_F[x:=t]}{A}{\Gamma}\}\\
        \interpaaux{b \Rightarrow A}{\Gamma} &\triangleq \left\{\omega \,|\, \interpbe{b}(\omega) \text{ is defined } \wedge \left(\interpbe{b}(\omega)=\fekeyword{true} \Rightarrow \omega \in \interpaaux{A}{\Gamma}\right)\right\}\\
        \interpaaux{b}{\Gamma} &\triangleq \{\omega \,|\, \interpbe{b}(\omega)=\fekeyword{true}\}\\
        \interpaaux{P}{\Gamma} &\triangleq \Gamma\\
        \interpaaux{\lfp{A}}{\Gamma} &\triangleq \LFP{S}{\interpaaux{A}{S}}\\
        \fvaaux{A_1 * A_2}{F} &\triangleq \fvaaux{A_1}{F} \cup \fvaaux{A_2}{F}\\
        \fvaaux{A_1 \vee A_2}{F} &\triangleq \fvaaux{A_1}{F} \cup \fvaaux{A_2}{F}\\
        \fvaaux{\existstp{x}{t} A}{F} &\triangleq \fvaaux{A}{F} \backslash \{x\}\\
        \fvaaux{b \Rightarrow A}{F} &\triangleq \fvbe(b) \cup \fvaaux{A}{F}\\
        \fvaaux{b}{F} &\triangleq \fvbe(b)\\
        \fvaaux{P}{F} &\triangleq F\\
        \fvaaux{\lfp{A}}{F} &\triangleq \LFP{S}{\fvaaux{A}{S}}
    \end{align*}
}
}

\revision{In the \isabelle{} formalization, we use de Bruijn indices, and prove the following overapproximation of $\fva(\lfp{A})$.
{
    \small
    \[
    \fva(\lfp{A}) \subseteq \bigcup_{x \in \fva(A)}\{0,1,\ldots,x\}
    \]
}
}

The extension of $\wfc$, $\fvc$, and $\wrc$ to all compositions of front-end statements is as follows.

{
    \small
    \begin{align*}
        \wfc^{\Delta_F}(\feskip) &\triangleq \text{true}\\
        \wfc^{\Delta_F}(c_1;c_2) &\triangleq \wfc^{\Delta_F}(c_1) \wedge \wfc^{\Delta_F}(c_2)\\
        \wfc^{\Delta_F}(\feif{b}{c_1}{c_2}) &\triangleq \wfc^{\Delta_F}(c_1) \wedge \wfc^{\Delta_F}(c_2)\\
        \wfc^{\Delta_F}(\fewhile{b}{I}{c}) &\triangleq \wfc^{\Delta_F}(c) \wedge \selfFraming{I} \wedge \stableUnder{b}{I} \revision{\wedge \wfa^{\Delta_F}(I)}\\
        \wfc^{\Delta_F}(\fepar{P_1}{c_1}{Q_1}{P_2}{c_2}{Q_2}) &\triangleq \wfc^{\Delta_F}(c_1) \wedge \wfc^{\Delta_F}(c_2) \wedge{} \\
        &\phantom{\triangleq} \selfFraming{P_1} \wedge \selfFraming{Q_1} \wedge{} \\
        &\phantom{\triangleq} \selfFraming{P_2} \wedge \selfFraming{Q_2} \wedge{} \\
        &\phantom{\triangleq} \wrc(c_1) \cap (\fvc(c_2) \cup \fva(P_2) \cup \fva(Q_2)) = \varnothing \wedge{} \\
        &\phantom{\triangleq} \wrc(c_2) \cap (\fvc(c_1) \cup \fva(P_1) \cup \fva(Q_1)) = \varnothing \revision{\wedge{}}\\
        &\phantom{\triangleq} \revision{\wfa^{\Delta_F}(P_1) \wedge \wfa^{\Delta_F}(Q_1) \wedge \wfa^{\Delta_F}(P_2) \wedge \wfa^{\Delta_F}(Q_2)}\\
        \fvc(\feskip) &\triangleq \varnothing\\
        \fvc(c_1;c_2) &\triangleq \fvc(c_1) \cup \fvc(c_2)\\
        \fvc(\feif{b}{c_1}{c_2}) &\triangleq \fvbe(b) \cup \fvc(c_1) \cup \fvc(c_2)\\
        \fvc(\fewhile{b}{I}{c}) &\triangleq \fvbe(b) \cup \fvc(c)\\
        \fvc(\fepar{P_1}{c_1}{Q_1}{P_2}{c_2}{Q_2}) &\triangleq \fvc(c_1) \cup \fvc(c_2)\\
        \wrc(\feskip) &\triangleq \varnothing\\
        \wrc(c_1;c_2) &\triangleq \wrc(c_1) \cup \wrc(c_2)\\
        \wrc(\feif{b}{c_1}{c_2}) &\triangleq \wrc(c_1) \cup \wrc(c_2)\\
        \wrc(\fewhile{b}{I}{c}) &\triangleq \wrc(c)\\
        \wrc(\fepar{P_1}{c_1}{Q_1}{P_2}{c_2}{Q_2}) &\triangleq \wrc(c_1) \cup \wrc(c_2)
    \end{align*}
}

The small-step operational semantic reduction rules and CSL rules for all the control structures are shown in \figref{fig:fe-stmt-opsem} and \figref{fig:fe-stmt-csl}, respectively.

\begin{figure}[ht]
    \footnotesize
    \begin{mathpar}
        \inferhref{RedPrim}{rule:RedPrim}
        {\sigma \stackrel{c}{\to}_{\Delta_F} \sigma'}
        {\fered{c}{\sigma}{\Delta_F}{\feskip}{\sigma'}}

        \and

        \inferhref{RedSeqSkip}{rule:RedSeqSkip}
        {}
        {\fered{\feskip;c}{\sigma}{\Delta_F}{c}{\sigma}}

        \and

        \inferhref{RedSeq}{rule:RedSeq}
        {\fered{c_1}{\sigma}{\Delta_F}{c_1'}{\sigma'}}
        {\fered{c_1;c_2}{\sigma}{\Delta_F}{c_1';c_2}{\sigma'}}

        \and

        \inferhref{RedIfT}{rule:RedIfT}
        {\interpbe{b}(\embeds(\sigma))=\fekeyword{true}}
        {\fered{\feif{b}{c_1}{c_2}}{\sigma}{\Delta_F}{c_1}{\sigma}}

        \and

        \inferhref{RedIfF}{rule:RedIfF}
        {\interpbe{b}(\embeds(\sigma))=\fekeyword{false}}
        {\fered{\feif{b}{c_1}{c_2}}{\sigma}{\Delta_F}{c_2}{\sigma}}

        \inferhref{RedLoop}{rule:RedLoop}
        {}
        {\fered{\fewhile{b}{I}{c}}{\sigma}{\Delta_F}{\feif{b}{c;\fewhile{b}{I}{c}}{\feskip}}{\sigma}}

        \and

        \inferhref{RedParSkip}{rule:RedParSkip}
        {}
        {\fered{\fepar{P_1}{\feskip}{Q_1}{P_2}{\feskip}{Q_2}}{\sigma}{\Delta_F}{\feskip}{\sigma}}

        \and

        \inferhref{RedParL}{rule:RedParL}
        {\fered{c_1}{\sigma}{\Delta_F}{c_1'}{\sigma'}}
        {\fered{\fepar{P_1}{c_1}{Q_1}{P_2}{c_2}{Q_2}}{\sigma}{\Delta_F}{\fepar{P_1}{c_1'}{Q_1}{P_2}{c_2}{Q_2}}{\sigma'}}

        \and

        \inferhref{RedParR}{rule:RedParR}
        {\fered{c_2}{\sigma}{\Delta_F}{c_2'}{\sigma'}}
        {\fered{\fepar{P_1}{c_1}{Q_1}{P_2}{c_2}{Q_2}}{\sigma}{\Delta_F}{\fepar{P_1}{c_1}{Q_1}{P_2}{c_2'}{Q_2}}{\sigma'}}
    \end{mathpar}
    \caption{Small-step operational semantic reduction rules of the form $\fered{c}{\sigma}{\Delta_F}{c'}{\sigma'}$ for front-end statements, where $c$ and $c'$ are statements, $\sigma$ and $\sigma'$ are program states, $\Delta_F$ is a front-end type context.}
    \label{fig:fe-stmt-opsem}
\end{figure}

\begin{figure}[ht]
    \footnotesize
    \begin{mathpar}
        \inferhref{RuleSkip}{rule:RuleSkip}
        {\selfFraming{P}}
        {\csl{\Delta_F}{P}{\feskip}{P}}

        \and

        \inferhref{RulePrim}{rule:RulePrim}
        {\selfFraming{P} \\
        \selfFraming{Q} \\
        \cslp{\Delta_F}{P}{c}{Q}}
        {\csl{\Delta_F}{P}{c}{Q}}

        \and

        \inferhref{RuleSeq}{rule:RuleSeq}
        {\csl{\Delta_F}{P}{c_1}{R} \\
        \csl{\Delta_F}{R}{c_2}{Q}}
        {\csl{\Delta_F}{P}{c_1;c_2}{Q}}

        \and

        \inferhref{RuleIf}{rule:RuleIf}
        {\framedBy{A}{b} \\
        \csl{\Delta_F}{A \cap b}{c_1}{B_1} \\
        \csl{\Delta_F}{A \cap \neg b}{c_2}{B_2}}
        {\csl{\Delta_F}{A}{\feif{b}{c_1}{c_2}}{B_1 \cup B_2}}

        \and

        \inferhref{RuleWhile}{rule:RuleWhile}
        {\framedBy{P}{b} \\
        \stableUnder{b}{P} \\
        \csl{\Delta_F}{P \cap b}{c}{P}}
        {\csl{\Delta_F}{P}{\fewhile{b}{I}{c}}{P \cap \neg b}}

        \and

        \inferhref{RulePar}{rule:RulePar}
        {\csl{\Delta_F}{P_1}{c_1}{Q_1} \\
        \csl{\Delta_F}{P_2}{c_2}{Q_2} \\
        \fvc(c_1) \cap \wrc(c_2) = \varnothing \\
        \fvc(c_2) \cap \wrc(c_1) = \varnothing \\
        \forall x \in \wrc(c_1).\, P_1 \text{ and } Q_1 \text{ are independent of } x \\
        \forall x \in \wrc(c_2).\, P_2 \text{ and } Q_2 \text{ are independent of } x}
        {\csl{\Delta_F}{P_1*P_2}{\fepar{A_1}{c_1}{B_1}{A_2}{c_2}{B_2}}{Q_1*Q_2}}

        \and

        \inferhref{RuleFrame}{rule:RuleFrame}
        {\csl{\Delta_F}{P}{c}{Q} \\
        \selfFraming{R} \\
        \forall x \in \wrc(c).\, R \text{ is independent of } x}
        {\csl{\Delta_F}{P*R}{c}{Q*R}}

        \and

        \inferhref{RuleConseq}{rule:RuleConseq}
        {\csl{\Delta_F}{P}{c}{Q} \\
        P' \subseteq P \\
        Q \subseteq Q' \\
        \selfFraming{P'} \\
        \selfFraming{Q'}}
        {\csl{\Delta_F}{P'}{c}{Q'}}



    \end{mathpar}
    \caption{CSL rules of the form $\csl{\Delta_F}{P}{c}{Q}$ for front-end statements, where $\Delta_F$ is a front-end type context, $c$ is a statement, $P$ and $Q$ are semantic assertions (\ie{} sets of front-end ghost states). The separating conjunction $P*Q$ of semantic assertions $P$ and $Q$ is defined as $\{\omega \,|\, \existsntp{\omega_1 \in P,\omega_2 \in Q} \omega=\omega_1 \oplus \omega_2\}$. A semantic assertion $P$ is independent of a local variable $x$ iff: $\forall \omega, v.\, \omega \in P \Leftrightarrow \omega[v/x] \in P$.}
    \label{fig:fe-stmt-csl}
\end{figure}

The full definition of $\tra{\cdot}$ is shown in \figref{fig:asst-tr}.

\begin{figure}
    \small
    \begin{align*}
        \tra{A_p} &\triangleq \trap{A_p} &
        \tra{b} &\triangleq \trbe{b} \\
        \tra{A_1 * A_2} &\triangleq \tra{A_1} * \tra{A_2} &
        \tra{A_1 \vee A_2} &\triangleq \tra{A_1} \vee \tra{A_2} \\
        \tra{b \Rightarrow A} &\triangleq \trbe{b} \Rightarrow \tra{A} &
        \tra{\existstp{x}{t} A} &\triangleq
        \existstp{x}{\trt{t}} \haveinv(t,x) * \tra{A} \\
        \revision{\tra{P}} &\mathrel{\revision{\triangleq}} \revision{P} &
        \revision{\tra{\lfp{A}}} &\mathrel{\revision{\triangleq}} \revision{\lfp{\tra{A}}}
    \end{align*}
    \caption{Extending the input assertion translation $\trap{\cdot}$ to a translation $\tra{\cdot}$ of all front-end assertions.}
    \label{fig:asst-tr}
\end{figure}

The full definition of $\trc{\cdot}$ is as follows, where $C_p \in C_{Fp}$ is a front-end primitive statement, and $\vhavoc{V}$ is a shorthand for $\vhavoc{x_1}; \cdots; \vhavoc{x_n}$ for finite set $V=\{x_1,\ldots,x_n\}$. Here we assume that the IVL supports (1) disjunction and negation of pure boolean expressions, and (2) directly using pure boolean expressions as assertions. It can therefore construct assertions including $\tra{I}*\left(\trbe{b} \vee \neg\trbe{b}\right)$ as in the translation of front-end loops.

{
    \footnotesize
    \begin{align*}
        \trc{\feskip} &\triangleq (\vskipp, \varnothing)\\
        \trc{C_p} &\triangleq \left(\trcp{C_p}, \varnothing\right)\\
        \trc{c_1;c_2} &\triangleq \left(\trcm{c_1};\trcm{c_2}, \trcext{c_1} \cup \trcext{c_2}\right)\\
        \trc{\feif{b}{c_1}{c_2}} &\triangleq \left(\vif{\trbe{b}}{\trcm{c_1}}{\trcm{c_2}}, \trcext{c_1} \cup \trcext{c_2}\right)\\
        \trc{\fewhile{b}{I}{c}} &\triangleq \left( \vexhale{\tra{I}*\left(\trbe{b} \vee \neg\trbe{b}\right)}; \vhavoc{\wrc(c)}; \vinhale{\tra{I}*\neg\trbe{b}}, \phantom{\trcm{c}}\right.\\
        &\phantom{\triangleq} \left. \left\{ \vinhale{\tra{I}*\tra{b}}; \trcm{c}; \vexhale{\tra{I}*\left(\trbe{b} \vee \neg\trbe{b}\right)} \right\} \cup \trcext{c} \right)\\
        \trc{\fepar{P_1}{c_1}{Q_1}{P_2}{c_2}{Q_2}} &\triangleq \left( \vexhale{\tra{P_1*P_2}}; \vhavoc{\wrc(c_1) \cup \wrc(c_2)}; \vinhale{\tra{Q_1*Q_2}}, \phantom{\trcm{c}}\right.\\
        &\phantom{\triangleq} \left\{ \vinhale{\tra{P_1}}; \trcm{c_1}; \vexhale{\tra{Q_1}} \right\} \cup \\
        &\phantom{\triangleq} \left. \left\{ \vinhale{\tra{P_2}}; \trcm{c_2}; \vexhale{\tra{Q_2}} \right\} \cup \trcext{c_1} \cup \trcext{c_2} \right)
    \end{align*}
}

The requirements for the input front-end components are listed in \reqref{req:fe}.

\begin{requirement}\label{req:fe}
    The following properties need to be satisfied by the input front-end components:
    \begin{itemize}
        \item $\interpbena$ is pure, \ie{} $\forall b \in BE_F, \omega \in \Sigma_F$, $\interpbe{b}(\omega)$ is defined iff $\interpbe{b}(\core{\omega})$ is defined, and $\interpbe{b}(\omega)=\interpbe{b}(\core{\omega})$ when they are both defined.
        \item For any boolean expression $b \in BE_F$, $\fvbe(b)$ overapproximates the free variables of $b$, \ie{} $\forall x \notin \fvbe(b),v \in V_F,\omega \in \Sigma_F$, $\interpbe{b}(\omega)$ is defined iff $\interpbe{b}(\omega[v/x])$ is defined, and $\interpbe{b}(\omega)=\interpbe{b}(\omega[v/x])$ when they are both defined.
        \item For any input assertion $A_p \in A_{Fp}$, $\fva(A_p)$ overapproximates the free variables of $A_p$, \ie{} $\forall x \notin \fva(A_p), \omega \in \Sigma_F, \revision{v \in V_F}$, $\omega \in \interpa{A_p}$ iff $\omega[v/x] \in \interpa{A_p}$.
        \item The written variables of well-formed primitive statements must be declared in the type context, \ie{} for any $c \in C_{Fp}$, if $\wfc^{\Delta_F}(c)$, then $\wrc(c) \subseteq \dom(\Delta_F)$.
        \item The CSL rules for any \revision{well-formed} primitive statement $C$ are affine, \ie{} if $\csl{\Delta_F}{P}{C}{Q}$ for some semantic assertions $P$ and $Q$, then $\csl{\Delta_F}{\monotonize{P}}{C}{\monotonize{Q}}$. Here, for any semantic assertion $A$, $\monotonize{A} \triangleq \{\omega \,|\, \existsntp{\omega_0} \omega \succeq \omega_0 \wedge \omega_0 \in A\}$ is the minimal semantic assertion that is monotone (a semantic assertion $P$ is monotone iff $\forall \omega \in P$ and $\omega' \succeq \omega$, $\omega' \in P$) and that subsumes $A$.
        \item The operational semantic reduction rules of any primitive statement $c \in C_{Fp}$ do not modify local variables outside $\wrc(c)$, \ie{} if $\sigma \stackrel{c}{\to}_{\Delta_F} \sigma'$, then $\forall x \notin \wrc(c)$, $\sigma(x)=\sigma'(x)$.
        \item For any primitive statement $c \in C_{Fp}$, any finite number of changes outside $\fvc(c)$ does not affect the reduction of $c$, \ie{} if two states $\sigma_1,\sigma_2$ agree on a set $X$ whose complement is finite and that subsumes $\fvc(c)$, and they have the same program heap, and $\sigma_1 \stackrel{c}{\to}_{\Delta_F} \sigma_1'$, then there exists $\sigma_2'$ such that $\sigma_2 \stackrel{c}{\to}_{\Delta_F} \sigma_2'$, and $\sigma_1'$ and $\sigma_2'$ agree on $X$ and the whole program heap.
        \item The operational semantic reduction of any \revision{well-formed} primitive statement preserves well-typedness, \ie{} if $\sigma \stackrel{c}{\to}_{\Delta_F} \sigma'$ for some $\sigma$ that is well-typed under $\Delta_F$, then $\sigma'$ is also well-typed under $\Delta_F$.
        \item The precondition of each CSL rule for well-formed primitive statements guarantees progression for well-typed states. Specifically, if $\cslp{\Delta_F}{P}{c}{Q}$ for some primitive statement $c \in C_{Fp}$ and monotone semantic assertion $P \subseteq \mathcal{P}(\Sigma_F)$, then for any program state $\sigma_p \in \Sigma_P$ that is well-typed \wrt{} $\Delta_F$, if $\embeds(\sigma_p)=\sigma \oplus \sigma_f$ such that $\sigma \in P$ and both $\sigma$ and $\sigma_f$ are stable, then there exists $\sigma_p' \in \Sigma_P$ such that $\sigma_p \stackrel{c}{\to}_{\Delta_F} \sigma_p'$.
        \item The front-end CSL rules for well-formed primitive statements are sound \wrt{} operational semantics: for any well-formed primitive statement, the post-state after the reduction satisfies the postcondition, whenever starting from a well-typed pre-state satisfying the precondition. Specifically, let $c \in C_{Fp}$ be well-formed and $\sigma_p$ be well-typed \wrt{} $\Delta_F$. If $\cslp{\Delta_F}{P}{c}{Q}$ for monotone semantic assertions $P$ and $Q$, $\embeds{}(\sigma_p)=\sigma \oplus \sigma_f$ such that $\sigma \in P$ and both $\sigma$ and $\sigma_f$ are stable, and $\sigma_p \stackrel{c}{\to}_{\Delta_F} \sigma_p'$, then there exist $\sigma'$ and $\sigma_f'$ such that $\sigma_f$ and $\sigma_f'$ agree on the ghost heap, $\sigma'$ is stable, $\embeds{}(\sigma_p')=\sigma' \oplus \sigma_f'$, and $\sigma' \in Q$.
    \end{itemize}
\end{requirement}

The requirements for the translation of front-end types and values are listed in \reqref{req:tp-val}.

\begin{requirement}\label{req:tp-val}
    For the translation of front-end types and values, $\trt{\cdot}$, $\trv{\cdot}$, and $\haveinv$ should satisfy the following properties:
    \begin{itemize}
        \item $\trt{\cdot}$ translates types correctly. Specifically, if $\tp(v)=t$ for any front-end value $v$ and type $t$, then $\trv{v}$ is of type $\trt{t}$.
        \item $\trv{\cdot}$ is injective. Specifically, $\trv{v_1}=\trv{v_2}$ implies $v_1=v_2$ for any front-end values $v_1$ and $v_2$.
        \item $\haveinv$ correctly reflects the bijectivity of the translation of values of all types. Specifically, if $\haveinv(t,x)$ is undefined for front-end type $t \in T_F$ for all $x$, then any value $v$ of type $\trt{t}$ in the IVL must correspond to some front-end value $v_f$ of type $t$ (\ie{} $\trv{v_f}=v$); if $\haveinv(t,x)=e$, then for any IVL state $\Omega \in \Sigma_V$ with $\Omega(x)=v$ for $v$ of type $\trt{t}$, $e(\Omega)=\vkeyword{true}$ iff there exists $v_f$ of type $t$ such that $\trv{v_f}=v$.
        \item The expressions $\haveinv$ returns are pure and monotone. Specifically, if $\haveinv(t,x)=e$, then for any $\Omega \in \Sigma_V$, $e(\Omega)$ is defined iff $e(\core{\Omega})$ is defined, and $e(\Omega)=e(\core{\Omega})$ when they are both defined; for any $\Omega,\Omega' \in \Sigma_V$, if $e(\Omega)$ is defined and $\Omega' \succeq \Omega$, then $e(\Omega')$ is also defined and evaluates to the same value as $e(\Omega)$ does.
    \end{itemize}
\end{requirement}

The requirement for the translation of primitive statements $\trcp{\cdot}$ is shown as \reqref{req:stmt}.

\begin{requirement}\label{req:stmt}
$\trcp{\cdot}$ is sound: it preserves the semantics of any primitive statement $C$. Specifically, if $\axiomSem[\trctxt{\Delta_F}]{P}{\trcp{C}}{Q}$ in the IVL for monotone IVL assertions $P$ and $Q$, then $\csl{\Delta_F}{\btr(P)}{C}{\btr(Q)}$ in the CSL of the front-end. Here, for any IVL semantic assertion $A$, $\btr(A) \triangleq \{\omega \,|\, \trs{\omega} \in A\}$ is the backward translation of $A$, \ie{} a collection of all front-end states whose encoded states satisfy $A$.
\end{requirement}

We prove \thmref{thm:csl-sound} in the framework, which states the soundness of the CSL rules in \figref{fig:fe-stmt-csl} \wrt{} the small-step operational semantics in \figref{fig:fe-stmt-opsem}.

\begin{theorem}[Soundness of the CSL in \figref{fig:fe-stmt-csl}]\label{thm:csl-sound}
    If $\csl{\Delta_F}{P}{C}{Q}$ for well-formed front-end statement $C$ and monotone $P$ and $Q$, then for any well-typed stable program state $\sigma$ such that $\embeds{}(\sigma) \in P$, executing $C$ from $\sigma$ never aborts, and for all $\sigma'$ that $\feredclo{C}{\sigma}{\Delta_F}{\feskip}{\sigma'}$, we have $\embeds{}(\sigma') \in Q$. Here $\feredclo{\cdot}{\cdot}{\cdot}{\cdot}{\cdot}$ is the reflexive transitive closure of $\fered{\cdot}{\cdot}{\cdot}{\cdot}{\cdot}$.
\end{theorem}

\section{Full Definitions and Semantics of the Instantiated Front-Ends}
\label{appsec:instantiation}

\subsection{Encoding Rational Permission Amounts into Real Permission Amounts}
\label{appsubsec:frac}

The interpretation of $\feaccpred{x}{f}{p}$ is defined in \figref{fig:frac-acc}.

{
    \begin{figure}[ht]
    \footnotesize
    \begin{mathpar}
        \inferhref{AccPos}{rule:AccPos}
        {\interpe{x}((s, (h, \pi)))=a \\
        a \text{ is a heap location identifier} \\
        \interpe{p} = p_v > 0 \\
        \pi(a.f) \ge p_v \\
        h(a.f) = v \\
        \Delta_F(f) = \tp{}(v)}
        {(s, (h, \pi)) \in \interpa{\feaccpred{x}{f}{p}}}

        \and

        \inferhref{AccZero}{rule:AccZero}
        {\interpe{x}((s, (h, \pi)))=r \\
        r = \fenull \text{ or } r \text{ is a heap location identifier} \\
        \interpe{p} = 0 \\
        \Delta_F(f) \text{ is defined}}
        {(s, (h, \pi)) \in \interpa{\feaccpred{x}{f}{p}}}
    \end{mathpar}
    \caption{The interpretation of $\feaccpred{x}{f}{p}$ in the rational permissioned front-end.}
    \label{fig:frac-acc}
    \end{figure}
}

\revision{
Every input assertion is well-formed in this front-end, \ie{} $\wfa^{\Delta_F}(A_p) \triangleq \text{true}$ for any input assertion $A_p$.
}

We can define syntactically the type $\tp{}_{\Delta_F}(e)$ for each expression $e$ by an induction on the structure of $e$. The definition guarantees that: for any expression $e$ of type $\tp_{\Delta_F}(e)=t$, if it evaluates to $v$ at some well-typed state $\omega$, then $v$ is of type $t$.
Using this definition, the well-formedness predicate for primitive statements is defined as follows.

{
    \small
    \begin{align*}
        \wfc^{\Delta_F}(\feassign{x}{e}) &\triangleq \existsntp{t} \Delta_F(x)=t \wedge \tp_{\Delta_F}(e)=t \\
        \wfc^{\Delta_F}(\fefieldassign{x}{f}{e}) &\triangleq \tp_{\Delta_F}(x)=\fekeyword{Ref} \wedge (\existsntp{t} \Delta_F(f)=t \wedge \tp_{\Delta_F}(e)=t) \\
        \wfc^{\Delta_F}(\fealloc{x}{f_1,\ldots,f_n}) &\triangleq \Delta_F(x)=\fekeyword{Ref} \wedge f_1,\ldots,f_n \text{ are distinct } \wedge \forallntp{i} \Delta_F(f_i) \text{ is defined} \\
        \wfc^{\Delta_F}(\fefree{x.f}) &\triangleq \top
    \end{align*}
}

The operational semantic reduction rules and CSL rules for the primitive statements are listed in \figref{fig:frac-opsem} and \figref{fig:frac-csl}, respectively.

\begin{figure}[ht]
        \footnotesize
    \begin{mathpar}
        \inferhref{RedAssignVar}{red:frac-assignvar}
        {e(\sigma)=v}
        {\sigma \xrightarrow{\feassign{x}{e}}_{\Delta_F} \sigma[v/x]}

        \and

        \inferhref{RedFree}{red:frac-free}
        {\sigma(x)=a \\
        \text{$a$ is a heap location identifier} \\
        \sigma(a.f) \ne \bot}
        {\sigma \xrightarrow{\fefree{x.f}}_{\Delta_F} \sigma[a.f := \bot]}

        \and

        \inferhref{RedAssignField}{red:frac-assignfield}
        {x(\sigma)=a \\
        \text{$a$ is a heap location identifier} \\
        e(\sigma)=v \\
        \tp{}(v) = \Delta_F(f) \\
        \sigma(a.f) \ne \bot}
        {\sigma \xrightarrow{\fefieldassign{x}{f}{e}}_{\Delta_F} \sigma[a.f := v]}

        \and

        \inferhref{RedAlloc}{red:frac-alloc}
        {\forallntp{f} (l.f \notin \dom(\sigma)) \\
        \Delta_F(x)=\fekeyword{Ref} \\
        \len{}(\vec{v}) = n \\
        \forallntp{0 < i \le n} \tp{}(\vec{v}[i])=\Delta_F(f_i)}
        {\sigma \xrightarrow{\fealloc{x}{f_1, \ldots, f_n}}_{\Delta_F} \sigma[l/x][l.f_1 := \vec{v}[1], \ldots, l.f_n := \vec{v}[n]]}
    \end{mathpar}
    \caption{Operational semantic reduction rules for primitive statements of the ractional permissioned front-end. We use $e(\sigma)$ to represent the evaluation of $e$ on program state $\sigma$ (whose result is the same as $\interpe{e}(\embeds{}(\sigma))$), $\sigma[v/x]$ to represent the state by substituting the value of local variable $x$ by value $v$ in $\sigma$, and $\sigma[a.f := \bot]$ or $\sigma[a.f := v]$ to represent the state by erasing the value at location $a.f$ or setting the value to $v$ in $\sigma$, respectively.}
    \label{fig:frac-opsem}
\end{figure}

\begin{figure}[ht]
    \footnotesize
    \begin{mathpar}
        \inferhref{RuleAssignVar}{rule:frac-assignvar}
        {\selfFraming{A} \\
        \framedBy{A}{e}}
        {\cslp{\Delta_F}{A}{\feassign{x}{e}}{A[e/x]}}

        \and

        \inferhref{RuleAssignField}{rule:frac-assignfield}
        {\selfFraming{A} \\
        A \Rightarrow \pointsto{x.f}{\_} \\
        \framedBy{A}{e}}
        {\cslp{\Delta_F}{A}{\fefieldassign{x}{f}{e}}{A[x.f:=e]}}

        \and

        \inferhref{RuleAlloc}{rule:frac-alloc}
        {}
        {\cslp{\Delta_F}{\top}{\fealloc{x}{f_1, \ldots, f_n}}{\feaccpred{x}{f_1}{1} * \cdots * \feaccpred{x}{f_n}{1}}}

        \and

        \inferhref{RuleFree}{rule:frac-free}
        {}
        {\cslp{\Delta_F}{\pointsto{x.f}{\_}}{\fefree{x.f}}{\top}}
    \end{mathpar}
    \caption{CSL rules for primitive statements of the ractional permissioned front-end. Here $\feaccpred{x}{f}{1}$ represents the typed version of $\pointsto{x.f}{\_}$, \ie{} while $\pointsto{x.f}{\_}$ asserts that the current state has full permission to the heap location $x.f$, $\feaccpred{x}{f}{1}$ further requires that the value $\omega(x.f)$ is well-typed \wrt{} the type context $\Delta_F$. We use $A[e/x]$ for the assertion $\{\omega[\interpe{e}(\omega)/x] \,|\, \omega \in A\}$, \ie{} updating the value of the local variable $x$ to the evaluation of $e$ for every state in $A$, and $A[x.f:=e]$ for the assertion $\{\omega[a.f := v] \,|\, \omega \in A, \interpe{x}(\omega)=a \text{ for some heap location identifier } a,\interpe{e}(\omega)=v\}$, \ie{} updating the value of the heap location $x(\omega).f$ to $e(\omega)$ for every $\omega \in A$ that $x$ and $e$ evaluate correctly.}
    \label{fig:frac-csl}
\end{figure}

The translation of all four primitive statements are listed as follows.

{
    \small
    \begin{align*}
        \trcp{\feassign{x}{e}} &\triangleq \vassign{x}{e}\\
        \trcp{\fefieldassign{x}{f}{e}} &\triangleq \vfieldassign{x}{f}{e} \\
        \trcp{\fealloc{x}{f_1,\ldots,f_n}} &\triangleq \vhavoc{x}; \vinhale{\vaccpred{x}{f_1}{1} * \cdots * \vaccpred{x}{f_n}{1}} \\
        \trcp{\fefree{x.f}} &\triangleq \vexhale{\vaccpred{x}{f}{1}}
    \end{align*}
}

\paragraph{\revision{A sound fragment of magic wands.}}
\revision{
The interpretation of $L \wand R$ is defined as follows:
{
    \small
\[
\interpa{L \wand R} \triangleq \left\{\omega \,|\, \forallntp{\omega_L \in \interpa{L}}{\omega_L \text{ is well-typed \wrt{} $\Delta_F$ and compatible with }\omega \Rightarrow \omega_L \oplus \omega \in \interpa{R}}\right\}
\]
}
}

\revision{
Well-formedness of $L \wand R$ is defined as $\wfa^{\Delta_F}(L \wand R) \triangleq \noratacc(L) \wedge \noratacc(R)$, where $\noratacc(A_b)$ if and only if $A_b$ does not access rational fields. Its concrete definition is given in \figref{fig:no-rat-acc}, where $\noratacce(e)$ of an expression $e$ holds if and only if the type of any field name $f$ in $e$ is not $\fekeyword{Rat}$ in $\Delta_F$.
}

\begin{figure}[ht]
    \small
    \revision{
    \begin{align*}
        \noratacc(\feaccpred{x}{f}{1}) &\triangleq \noratacce(x)\\
        \noratacc(e) &\triangleq \noratacce(e)\\
        \noratacc(A_1 * A_2) &\triangleq \noratacc(A_1) \wedge \noratacc(A_2)\\
        \noratacc(A_1 \vee A_2) &\triangleq \noratacc(A_1) \wedge \noratacc(A_2)\\
        \noratacc(e \Rightarrow A) &\triangleq \noratacce(e) \wedge \noratacc(A)\\
        \noratacc(\existstp{x}{t}{A}) &\triangleq \noratacc(A)
    \end{align*}
    }
    \caption{\revision{Definition of $\noratacc{}$.}}
    \label{fig:no-rat-acc}
\end{figure}

\subsection{Encoding Lazy Field Creation}
\label{appsubsec:dfc}

$\interpana$ for the three input assertions is shown in \figref{fig:dfc-asst}.
\revision{
Every input assertion is well-formed in this front-end, \ie{} $\wfa^{\Delta_F}(A_p) \triangleq \text{true}$ for any input assertion $A_p$.
}

\begin{figure}[ht]
    \footnotesize
    \begin{mathpar}
        \inferhref{DFCAcc}{rule:dfc-acc}
        {\interpe{x}((s, (h,\pi))) = a \\
        \text{$a$ is a heap location identifier} \\
        \interpe{p}((s, (h, \pi))) = p_v > 0 \\
        \pi(a.f) \ge p_v \\
        h(a.f) = \fenone \vee (h(a.f) = v \wedge \Delta_F(f) = \tp{}(v))}
        {\interpa{\feaccpred{x}{f}{p}} = 1}

        \and

        \inferhref{DFCAdded}{rule:dfc-init}
        {\interpe{x}((s, (h, \pi))) = a \\
        \text{$a$ is a heap location identifier} \\
        h(a.f) = v \ne \fenone}
        {\interpa{\initpred{x}{f}} = 1}

        \and

        \inferhref{DFCMissing}{rule:dfc-uninit}
        {\interpe{x}((s, (h, \pi))) = a \\
        \text{$a$ is a heap location identifier} \\
        h(a.f) = \fenone}
        {\interpa{\uninitpred{x}{f}} = 1}
    \end{mathpar}
    \caption{The assertion interpretation $\interpana$ for input assertions of the lazy field creation front-end.}
    \label{fig:dfc-asst}
\end{figure}

The operational semantic reduction rule for allocations is shown in \figref{fig:dfc-opsem}.

\begin{figure}[ht]
    \footnotesize
    \begin{mathpar}
        \inferhref{RedAlloc}{red:dfc-alloc}
        {\forall f. (l, f) \notin \dom(\omega) \\
        \Delta_F(x)=\fekeyword{Ref}}
        {\omega \xrightarrow{\fealloc{x}{f_1, \ldots, f_n}}_{\Delta_F} \omega[l/x][l.f_i := \fenone]}
    \end{mathpar}
    \caption{The operational semantic reduction rule for $\fealloc{x}{f_1,\ldots,f_n}$ in the lazy field creation front-end.}
    \label{fig:dfc-opsem}
\end{figure}

The translation of input assertions is defined as follows:

{
    \small
    \begin{align*}
        \trap{\feaccpred{x}{f}{p}} &\triangleq \tre{p}>0*\vaccpred{\tre{x}}{\addedf{}}{\tre{p}}*(\tre{x}.\addedf{}=\vtrue \Rightarrow \vaccpred{\tre{x}}{\ivlf{}}{\tre{p}})\\
        \trap{\initpred{x}{f}} &\triangleq \tre{x}.\addedf{}=\vtrue\\
        \trap{\uninitpred{x}{f}} &\triangleq \tre{x}.\addedf{}=\vfalse
    \end{align*}
}

The translation of primitive statements is defined as follows.

{
    \small
    \begin{align*}
        \trcp{\feassign{x}{e}} &\triangleq \vassign{x}{\tre{e}} \\
        \trcp{\fealloc{x}{f_1,\ldots,f_n}} &\triangleq \vhavoc{x}; \vinhale{\tra{\feaccpred{x}{f_1}{1}} * \cdots * \tra{\feaccpred{x}{f_n}{1}}} \\
        \trcp{\fefree{x}{f}} &\triangleq \vexhale{\tra{\feaccpred{x}{f}{1}}} \\
        \trcp{\fefieldassign{x}{f}{v}} &\triangleq \vif{\tre{x}.\addedf{}=\vfalse}{\vinhale{\vaccpred{\tre{x}}{\ivlf{}}{1}}}{\vskipp};\\
        &\phantom{\triangleq} \vfieldassign{\tre{x}}{\addedf{}}{\vtrue};
        \vfieldassign{\tre{x}}{\ivlf{}}{v}
    \end{align*}
}

\subsection{Encoding Immutable Heaps}
\label{appsubsec:immutable-heap}

The interpretation function $\interpana$ on input assertion $\iminitpred{x}{f}$ is shown in \figref{fig:im-asst}.
\revision{
Every input assertion is well-formed in this front-end, \ie{} $\wfa^{\Delta_F}(A_p) \triangleq \text{true}$ for any input assertion $A_p$.
}

\begin{figure}[ht]
    \footnotesize
    \begin{mathpar}
        \inferhref{ImInit}{rule:im-init-pred}
        {\interpe{x}(s, (h, \pi, h_{im})) = a \\
        a \text{ is a heap location identifier} \\
        h_{im}(\imacc{a}{f}) = v \\
        \tp{}(v) = \Delta_F^{im}(f)}
        {(s, (h, \pi, h_{im})) \in \interpa{\iminitpred{x}{f}}}
    \end{mathpar}
    \caption{The interpretation of $\iminitpred{x}{f}$ in the immutable heap front-end. We allow the regular heap and the immutable heap to have different type contexts, and use $\Delta_F^{im}$ to denote the type context of the immutable heap.}
    \label{fig:im-asst}
\end{figure}

The free variable set, written variable set, and well-formedness predicate for $\feinit{x}{f}{v}$ are naturally defined as follows.

{
    \small
    \begin{align*}
        \fvc{}(\feinit{x}{f}{v}) &\triangleq \{x,v\} \\
        \wrc{}(\feinit{x}{f}{v}) &\triangleq \{x\} \\
        \wfc^{\Delta_F}(\feinit{x}{f}{v}) &\triangleq x \ne v \wedge \Delta_F(x) = \fekeyword{Ref} \wedge \Delta_F(v) = \Delta_F(f) \ne \bot
    \end{align*}
}

The CSL rule and operational semantic reduction rule for $\feinit{x}{f}{v}$ are shown in \figref{fig:im-csl-opsem}.

\begin{figure}[ht]
    \footnotesize
    \begin{mathpar}
        \inferhref{RuleInit}{rule:im-init}
        {}
        {\csl{\Delta_F}{\top}{\feinit{x}{f}{v}}{\imacc{x}{f}=v}}

        \and

        \inferhref{RedInit}{red:im-init}
        {\forall f.\, \omega(\imacc{l}{f}) = \bot \\
        \Delta_F(x)=\fekeyword{Ref} \\
        \omega(v) = v_0}
        {\omega \xrightarrow{\feinit{x}{f}{v}}_{\Delta_F} \omega[l/x][\imacc{l}{f} := v_0]}
    \end{mathpar}
    \caption{The CSL rule and operational semantic reduction rule for the primitive statement $\feinit{x}{f}{v}$ in the immutable heap front-end.}
    \label{fig:im-csl-opsem}
\end{figure}

The translation of primitive statements is listed as follows.

{
    \small
    \begin{align*}
        \trcp{\feassign{x}{e}} &\triangleq \vassign{x}{\tre{e}} \\
        \trcp{\fefieldassign{x}{f}{e}} &\triangleq \vfieldassign{\tre{x}}{f_h(f)}{\tre{e}}\\
        \trcp{\fealloc{x}{f_1,\ldots,f_n}} &\triangleq \vhavoc{x}; \vinhale{\vaccpred{x}{f_h(f_1)}{1} * \cdots * \vaccpred{x}{f_h(f_n)}{1}} \\
        \trcp{\fefree{x.f}} &\triangleq \vexhale{\vaccpred{x}{f_h(f)}{1}} \\
        \trcp{\feinit{x}{f}{v}} &\triangleq \vhavoc{x}; \vinhale{\vaccpred{x}{f_v(f)}{\_} * x.f_v(f)=v}
    \end{align*}
}

\fi

\end{document}